\documentclass[prd,12pt,a4paper,nofootinbib,tightenlines,showpacs,showkeys]{revtex4}
\usepackage[left=2.5cm,right=2.5cm,top=2.5cm,bottom=2.5cm]{geometry}

\usepackage[frenchb,english]{babel}
\usepackage[latin1]{inputenc}
\usepackage{enumerate}
\usepackage{amsmath}
\usepackage{amsthm}
\usepackage{amssymb}
\usepackage{graphicx}
\usepackage{floatflt}
\usepackage{fancybox}
\usepackage{appendix}
\usepackage{enumitem}
\usepackage[lofdepth,lotdepth]{subfig}
\usepackage{array}

\usepackage{hyperref}
\usepackage[b]{esvect}

\usepackage{color}

\renewcommand{\Re}{\textrm{Re}}
\renewcommand{\Im}{\textrm{Im}}

\newcommand{\om}{\omega}

\newcommand{\eff}{{\rm eff}}

\newcommand{\m}{{\rm min}}
\newcommand{\M}{{\rm max}}

\renewcommand{\u}{{A-}} %acoustic mode
\renewcommand{\v}{{A+}} %acoustic mode
\newcommand{\+}{{S}} %dispersive mode
\renewcommand{\-}{{N}} %dispersive mode NEW
\newcommand{\s}{{B}} %boundary layer mode

\renewcommand{\vec}[1]{\mathbf{#1}} %or \vv
\newcommand{\p}{\partial}
\renewcommand{\d}{\mathrm{d}}

\newcommand{\be} {\begin{equation}}
\newcommand{\ee} {\end{equation}}
\newcommand{\bsub}{\begin{subequations}}
\newcommand{\esub}{\end{subequations}}
\newcommand{\bea}{\begin{eqnarray}}
\newcommand{\eea}{\end{eqnarray}}
\newcommand{\bi} {\begin{itemize}}
\newcommand{\ei} {\end{itemize}}
\newcommand{\ben} {\begin{enumerate}}
\newcommand{\een} {\end{enumerate}}
\newcommand{\bmat} {\begin{pmatrix}}
\newcommand{\emat} {\end{pmatrix}}
\newcommand{\bal} {\begin{aligned}}
\newcommand{\eal} {\end{aligned}}
\newcommand{\btab}{\begin{tabular}}
\newcommand{\etab}{\end{tabular}}

\newcommand{\eq}[1]{equation~\eqref{#1}}

\begin{document}
\selectlanguage{english}

\title{Slow sound laser in lined flow ducts}

\author{Antonin Coutant}
\email{antonin.coutant@univ-lemans.fr}
\affiliation{Laboratoire d'Acoustique de l'Université du Maine, Unite Mixte de Recherche 6613, Centre National de la Recherche Scientifique, Avenue O. Messiaen, F-72085 LE MANS Cedex 9, France}

\author{Yves Aurégan}
\email{yves.auregan@univ-lemans.fr}
\affiliation{Laboratoire d'Acoustique de l'Université du Maine, Unite Mixte de Recherche 6613, Centre National de la Recherche Scientifique, Avenue O. Messiaen, F-72085 LE MANS Cedex 9, France}

\author{Vincent Pagneux}
\email{vincent.pagneux@univ-lemans.fr}
\affiliation{Laboratoire d'Acoustique de l'Université du Maine, Unite Mixte de Recherche 6613, Centre National de la Recherche Scientifique, Avenue O. Messiaen, F-72085 LE MANS Cedex 9, France}

\date{\today}

\begin{abstract}
This work considers the propagation of sound in a waveguide with an impedance wall. In the low frequency regime, the first effect of the impedance is to decrease the propagation speed of acoustic waves. Therefore, a flow in the duct can exceed the wave propagation speed at low Mach numbers, making it effectively supersonic. This work analyzes a setup where the impedance along the wall varies such that the duct is supersonic then subsonic in a finite region and supersonic again. In this specific configuration, the subsonic region act as a resonant cavity, and triggers a laser-like instability. This work shows that the instability is highly subwavelength. Besides, if the subsonic region is small enough, the instability is static. This worl also analyzes the effect of a shear flow layer near the impedance wall. Although its presence significantly alter the instability, its main properties are maintained. This work points out the analogy between the present instability and a similar one in fluid analogues of black holes known as the black hole laser. 
\end{abstract}



\maketitle

%%%%%%%%%%%%%%%%%%%%%%%%%%%%%%%%%%%%%%%%%%%%%%%%%%%
%INTRODUCTION
%%%%%%%%%%%%%%%%%%%%%%%%%%%%%%%%%%%%%%%%%%%%%%%%%%%
\section{Introduction}

Acoustic liners in waveguides offer the interesting possibility to slow down sound waves. In a fluid, the speed of sound $c_0$ is controlled by both the density of the fluid and its stiffness (the adiabatic bulk modulus). Acoustic liners can be created using tubes mounted flush to the wall of the guide, which lower the effective stiffness of the medium, thereby decreasing the effective propagation speed $c_\eff$ of sound. This allows one to control and manipulate sound waves in guides. In particular, adding a flow of mean velocity $U_0$, an effective supersonic configuration ($c_\eff<U_0$) can be obtained in the duct at low Mach number ($M_0 = U_0/c_0 < 1$). Transsonic configurations, gradually varying from subsonic to supersonic, lead to a rich wave phenomenology~\cite{Auregan15,Auregan15b} such as amplification and highly non-reciprocal propagation and can even lead to instabilities. 
\bigskip

The existence of instabilities above acoustic materials in the presence of a grazing flow has been experimentally proven \cite{Ronneberger, Auregan2008, Marx}. Sometimes, it is difficult in computations to distinguish between real and numerical instabilities \cite{li2006time, burak2009validation, gabard2014full, xin2016numerical, pascal2017global, sebastian2017numerical}. An analysis of the different types of instability that can occur above a material is therefore of importance for a better understanding of the results of experiments or computations.  Transonic configurations allow modes with very different wavelengths to interact through the flow and instabilities can occur in cavities created by impedance changes, even if their sizes are very small compared to the acoustic wavelength.
\bigskip

\begin{figure}[!ht]
\centering
\includegraphics[width=0.6\columnwidth]{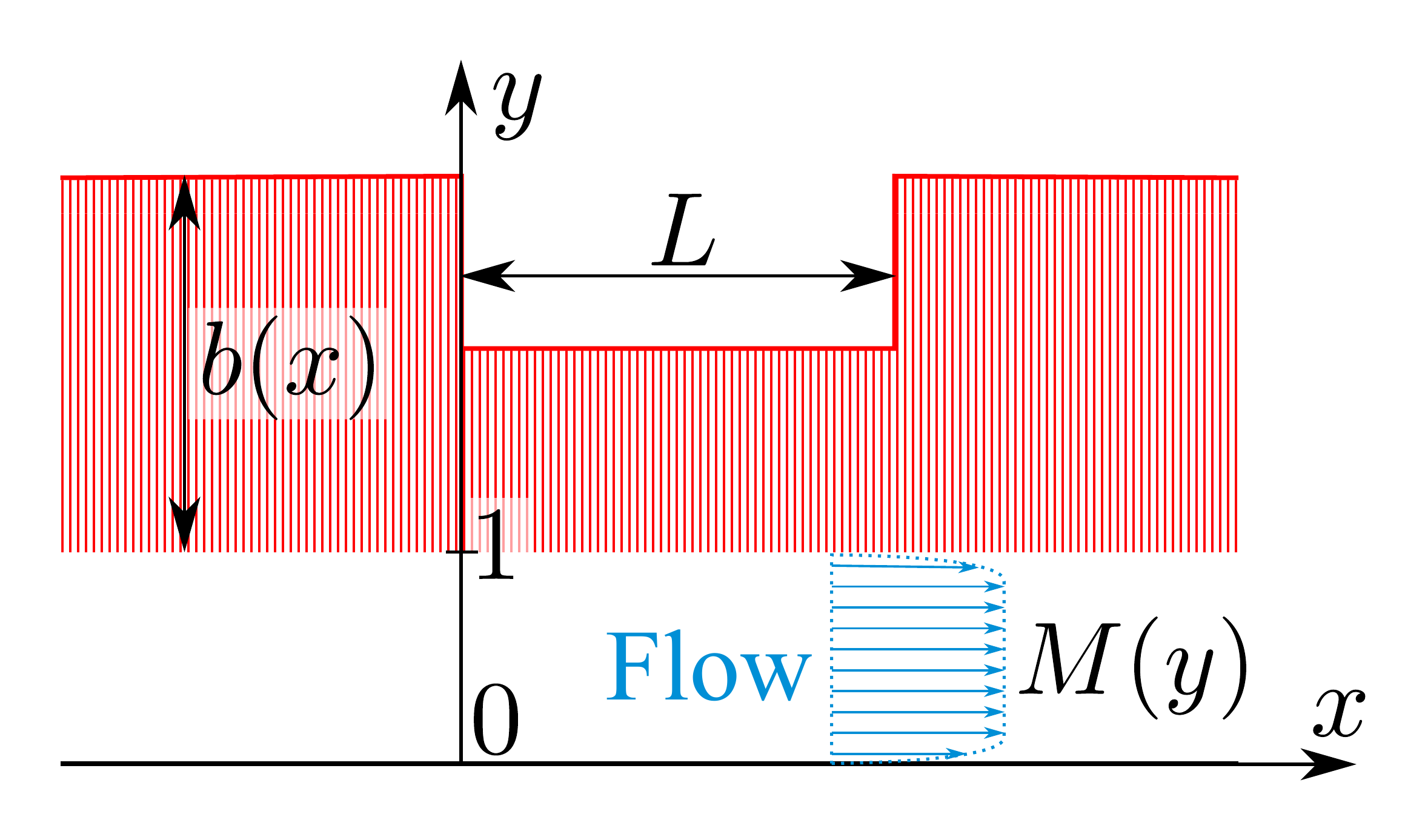}
\caption{Configuration of the problem. 
}
\label{Config_Fig} 
\end{figure}

In the absence of flow, two acoustic modes can propagate in the duct at low frequencies. In subsonic lined ducts, there are two additional modes, which are referred to as hydrodynamic modes~\cite{Rienstra03}. Moreover, one of them is a \emph{negative energy wave}. This means that its excitation lowers the total energy of system compared to the mean flow alone~\cite{Cairns79}. Coupling to this negative energy wave through a change in the impedance wall can therefore lead to amplification~\cite{Auregan15,Auregan15b}. In this work, we study a configuration consisting in a double transition: from supersonic to subsonic to supersonic again, see Fig.~\ref{Config_Fig}. In such a configuration, the negative energy wave couples to a resonant cavity (formed by the subsonic region), thereby generating an exponentially growing instability. A peculiarity of the obtained instability, is that it occurs at a much lower frequency than the natural frequencies of the system, such as the frequency associated with the quarter wavelength of the tubes or with the size of the cavity formed by the two transitions. 
\bigskip

A peculiar feature of transsonic flows has attracted a lot of attention in the last decade: they can provide a laboratory analogue of a black hole~\cite{Unruh81}. Interestingly, the instability studied in this work is closely related to the analogue of the Hawking radiation of black hole, and can be seen as a self-amplification of that radiation. This self-amplification is called the ``black hole laser'' in the analogue gravity community, and was studied in various contexts~\cite{Coutant10,Steinhauer14}. The analogue Hawking effect has lead to many experiments in the last decade, in media as diverse as water waves~\cite{Weinfurtner10,Euve15,Coutant17}, nonlinear optics~\cite{Drori18} or Bose-Einstein condensates~\cite{Steinhauer15,deNova18}, but despite many promising results, the full demonstration of the analogue Hawking radiation and its properties has not been achieved yet. Slow sound offers a promising system for its realization. 
\bigskip 

To describe this instability and analyze its main features, we use an effective one-dimensional model that was previously derived in~\cite{Auregan15,Auregan15b}. Moreover, we shall take into account the effect of a shear flow boundary layer near the impedance wall. In a majority of works, the boundary layer is taken to be infinitely thin, leading to the so-called Ingard-Myers boundary condition at the impedance wall~\cite{Ingard59,Myers80}. This boundary condition has however shown to be problematic, both from the theory~\cite{Brambley09} and experimental point of view~\cite{Renou11}. In this work, we will use an improved boundary condition, based on the work of Brambley~\cite{Brambley11,Brambley13}. When using the Ingard-Myers condition, unstable modes can be divided into two categories: static instabilities (purely imaginary frequency, hence non oscillatory) and dynamical instabilities (non zero real part). With the inclusion of the boundary layer correction in the improved boundary condition, static instabilities acquire a non-zero real part, proportional to the ratio of the thickness of the boundary layer with the duct width. 
\bigskip

The plan of the paper is as follows. In section II we present the model with the improved boundary condition. Then, we employ a Hamiltonian formalism to identify conserved quantities. In section III we discuss the spectrum of complex eigenfrequencies and separate the discussion of an infinitely thin boundary layer with a finite one. We then present our main conclusions.

\section{Propagation in a lined duct with a flow having a thin boundary layer}
%%%%%%%%%%%%%%%%%%%%%%%%%%%%%%%%%%%%%%%%%%%%%%%%%%%
%SLOW SOUND
%%%%%%%%%%%%%%%%%%%%%%%%%%%%%%%%%%%%%%%%%%%%%%%%%%%
We consider the propagation of sound waves in a straight two-dimensional duct with a uniform flow $U_0$ (see Fig.~\ref{Config_Fig}). In the following we work with adimensionalized quantities, using the speed of sound $c_0$, the density $\rho_0$ and the height of the duct $H$, all three are assumed to be constant. 
This means that the flow velocity is replaced by the (uniform) Mach number $M_0=U_0/c_0$. Inside the duct, the acoustic velocity field is potential $\vec v = \vec \nabla \phi$, and obeys the wave equation 
\be \label{Bulk_wave_eq}
D_t^2 \phi - \Delta \phi = 0, 
\ee
where $D_t = \p_t + M_0 \p_x$ is the convective derivative. The lower wall ($y=0$) of the waveguide is assumed to be hard. This means that the boundary condition is simply given by a vanishing transverse velocity, that is 
\be
(\p_y \phi)_{y=0} = 0. 
\ee
The upper wall ($y=1$) is compliant, and its boundary condition is defined by an impedance: the ratio of the pressure over the transverse acoustic velocity. In the absence of flow in the duct, the normalized impedance reads 
\be
Z_b = \frac{i}{\sigma \tan(b \om)}, 
\ee
where $b$ is the length of the mounted flush tubes, $\sigma$ is the percentage of open area, and $\omega$ the angular frequency. We now work in the low frequency regime, that is when $\om$ is small compared to the tube resonance frequency $\pi/2b$. In this regime, $\tan(b\, \om) \sim b\,\om$, and the compliant wall act as a spring of stiffness $\sigma b$. Without loss of generality, we will now assume that $\sigma = 1$. For a nonzero flow inside the duct, although we assume a constant profile (independent of $y$), we must take into account the thin boundary layer (thickness $\delta$) within which the Mach number decreases from its maximum value $M_0$ to zero. 
In a sequence of works~\cite{Brambley11,Brambley13}, Brambley derived general boundary conditions that takes into account the boundary layer at order $O(\delta)$. 
For this work in the low frequency limit ($\om \ll \pi/2b$), we will use an approximate boundary condition at order $O(\delta)$. For a wave of frequency $\om$ and wave number $k$, that is $\phi = \Re [\bar \phi(y) e^{-i \om t + i k x}]$, it is given by 
\be \label{Bram_BC}
(\p_y \phi)_{y=1} = \left(b(\om - M_0 k)^2 + \frac{\delta M_0 k^3}{\om}\right)\phi_{y=1}. 
\ee
In Appendix~\ref{BC_App}, we provide details on how it has been obtained and discuss its regime of validity. Notice that by taking $\delta = 0$, one recovers the standard Ingard-Myers condition.

%%%%%%%%%%%%%%%%%%%%%%%%%%%%%%%%%%%%%%%%%%%%%%%%%%%
%1D MODEL
%%%%%%%%%%%%%%%%%%%%%%%%%%%%%%%%%%%%%%%%%%%%%%%%%%%

To understand the propagation of waves in a duct of varying impedance, an effective one-dimensional model was derived in~\cite{Auregan15,Auregan15b}. The effective model is obtained by assuming a short transverse size of the duct compared with relevant wavelengths. As shown in~\cite{Auregan15,Auregan15b}, this allows us to obtain a simple model that displays the same number of propagating modes with the same properties (such as the sign of their energy). However, the model of~\cite{Auregan15,Auregan15b} used the Ingard-Myers boundary condition. Here we shall consider instead the more general condition \eqref{Bram_BC}. 

\subsection{Dispersion relation and propagating modes in the 1D model}
\begin{figure}[!ht]
\centering
\subfloat[]{\includegraphics[width=0.5\textwidth]{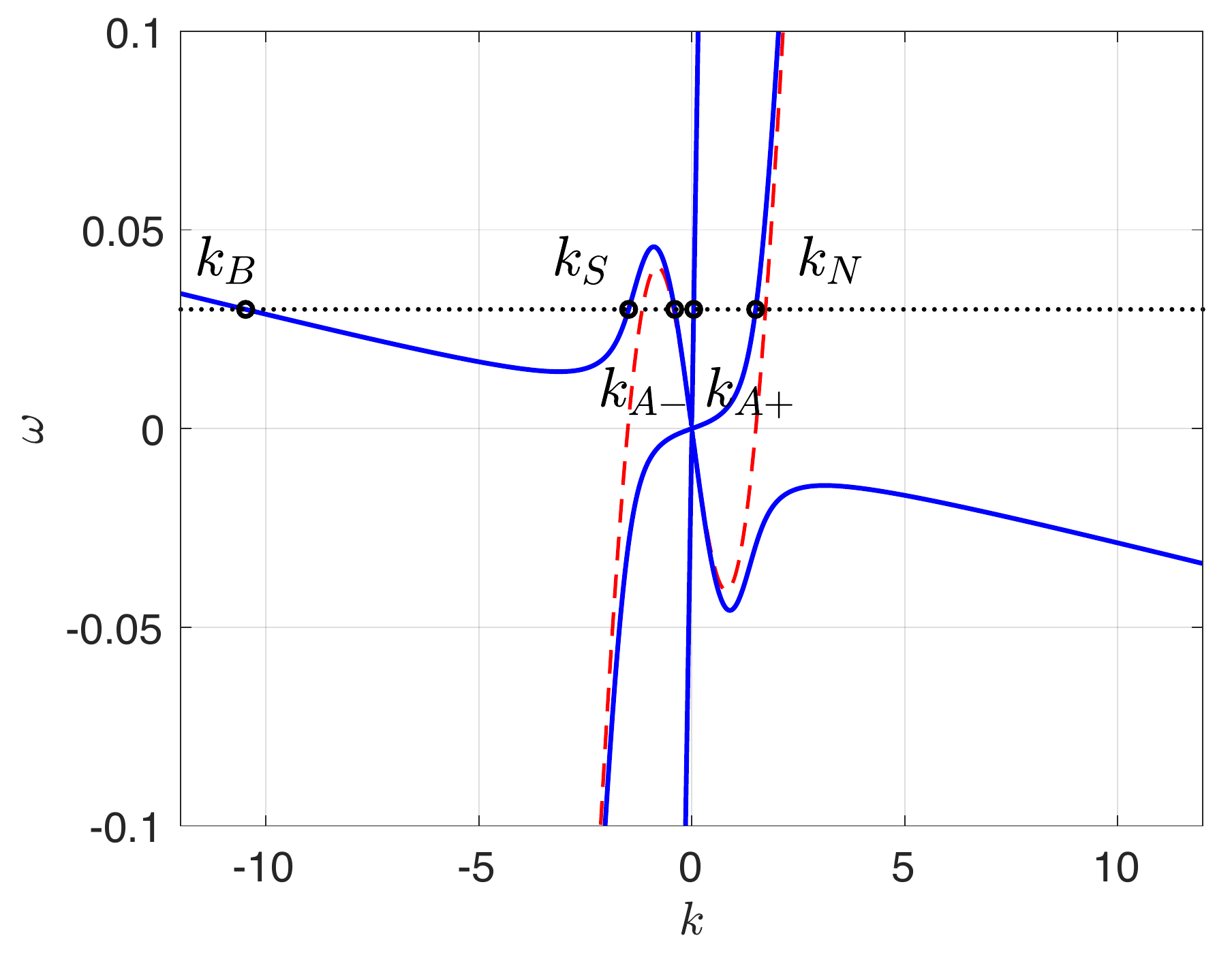} \label{1DeffDisp_Sub_Fig} }
\subfloat[]{\includegraphics[width=0.5\textwidth]{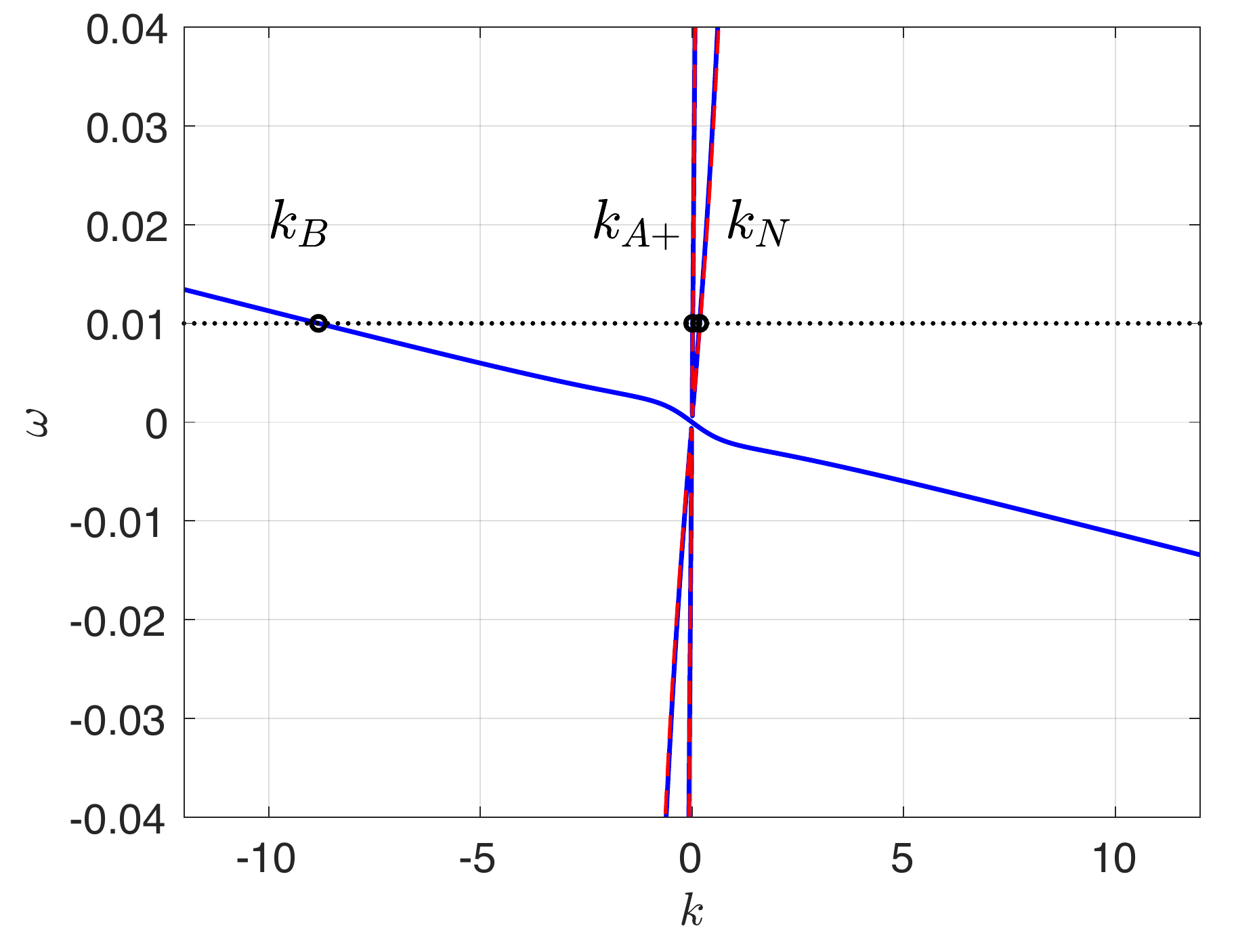} \label{1DeffDisp_Super_Fig} }
\caption{Dispersion relation obtained in the one-dimensional model, i.e. equation \eqref{1D_Disp_rel}. The solid blue line is the dispersion relation with the finite boundary layer correction \eqref{Bram_BC}, and the dashed red line is the dispersion relation in the Ingard-Myers limit ($\delta=0$). We have used $M_0=0.3$ and $\delta=0.005$, and $b=6$ on the left and $b=15$ on the right. [Note that in black and white, the curves $k_{A+}$ and $k_N$ with the two boundary conditions on the right side are indistinguishable.] 
}
\label{1DeffDisp_Fig} 
\end{figure}

We first quickly summarize how the effective model is obtained. First we define 
\bsub \bea
\varphi &=& \int_0^1 \phi(t,x,y) \d y, \\
\psi &=& \phi(t,x,1) .
\eea \esub
Integrating the wave equation \eqref{Bulk_wave_eq} across the duct, we obtain 
\be \label{Integrated_wave_eq}
D_t^2 \varphi - \p_x^2 \varphi - (\p_y \phi)_{y=1} = 0. 
\ee
We now assume that the transverse size of the duct is small enough so that the transverse dependence of the pressure field can be treated in a parabolic approximation, i.e. $\phi(t,x,y) = \phi_1(t,x) + y^2 \phi_2(t,x)$. In this approximation, the transverse derivative on the compliant wall can be written as a sum of $\varphi$ and $\psi$, indeed: 
\be \label{Parab_Approx}
\p_y \phi(t,x,1) = 3 \psi - 3 \varphi. 
\ee
We can now use this in equations \eqref{Integrated_wave_eq} and \eqref{Bram_BC}. For a single frequency and wave number mode, this leads to the dispersion relation 
\be \label{1D_Disp_rel}
\left(-(\om - M_0 k)^2 + k^2 + 3 \right) \left( b(\om - M_0 k)^2 + \frac{\delta M_0 k^3}{\om} - 3 \right) + 9 = 0. 
\ee
In Fig.~\ref{1DeffDisp_Fig} we represent the dispersion relation, and compare the Ingard-Myers condition with our modified condition \eqref{Bram_BC}. We now discuss the different propagating modes at a given frequency. 
To start we consider the long wavelength limit $k \to 0$ in the absence of flow ($M_0=0$), in which case the dispersion relation becomes 
\be \label{NoFlow_1D_Disp_Rel}
\om^2 = \frac{k^2}{1+b}. 
\ee
This means that long wavelength waves have a propagation speed modified by the impedance wall of $c_\eff = 1/\sqrt{1+b}$. Since this value is always lower than the speed of sound in free air, these waves are called ``slow sound waves''. Based on this, we distinguish two types of flow: effective subsonic ones, when $M_0\sqrt{1+b} < 1$, and effective supersonic ones, when $M_0\sqrt{1+b} > 1$. We will now discuss the various propagating modes for both types of flows, separating the case of a vanishing boundary layer (Ingard-Myers condition) and a nonzero one (boundary condition \eqref{Bram_BC}). 

Let us start by discussing the Ingard-Myers condition. For subsonic flows, see Fig.~\ref{1DeffDisp_Sub_Fig}, there are two frequency ranges separated by a threshold frequency $\om_\M$. When $\om < \om_\M$ there are four solutions. Two of them have long wavelength, and corresponds to acoustic modes $k_{A\pm}$ Doppler shifted by the flow. The two others are highly dispersive: their group velocity differs significantly from their phase velocity. These two extra modes would not exist in the absence of flow, and for that reason, are referred to as ``hydrodynamic modes''~\cite{Rienstra03}. A peculiarity of these new modes is that while one (noted $k_S$) has a positive energy, the other carries (noted $k_N$) a negative energy. For higher frequencies, $\om > \om_\M$, there are only two modes left, one with negative energy, and both propagating with the flow. In effective supersonic flows, see Fig.~\ref{1DeffDisp_Super_Fig}, there are two propagating modes for all frequencies, both propagating in the direction of the flow. Again, while one has a positive energy, the other has a negative energy. 

When using the improved boundary condition of \eqref{Bram_BC}, a new propagating mode appear. More precisely, there is a new threshold frequency $\om_\m < \om_\M$. In the range $\om_\m < \om < \om_\M$, the discussion is similar as before, but with a fifth mode, noted $k_\s$. This modes propagate very slowly (its group velocity being proportional to $\delta$) against the flow. when $\om=\om_\m$, this mode merge with $k_\+$ to give two evanescent modes. For $\om < \om_\m$ there is therefore only three propagating modes. Although this mode carries a positive energy, at low frequencies it couples to the other ones and alter significantly the laser effect. The asymptotic values when $\om_\m \ll \om \ll \om_\M$ of the different wavenumbers are given in Appendix~\ref{Roots_App}. 

In the rest of this work, we are interested in the effects of a change of the wall compliance along the duct. In other words, we will assume that $b$ is a function of $x$. Such inhomogeneity will couple together different modes sharing the same frequency. Due to the presence of negative energy waves, this can lead to amplification phenomena. In addition, when the change of wall compliance create a resonant cavity, this leads to a temporal instability. Such an instability is the acoustic analogue of the ``black hole laser'' that have been studied in various analogue models of gravity~\cite{Coutant10,Finazzi10,Michel13}.

%%%%%%%%%%%%%%%%%%%%%%%%%%%%%%%%%%%%%%%%%%%%%%%%%%%
%HAMILTONIAN FORMALISM OF 1D MODEL
%%%%%%%%%%%%%%%%%%%%%%%%%%%%%%%%%%%%%%%%%%%%%%%%%%%
\subsection{Conservation laws}
The existence of a conserved energy in ducts is \emph{a priori} non trivial, even for a conservative impedance as in our case (see e.g. the discussion of Möhring~\cite{Mohring01}). The question becomes increasingly delicate in the presence of a shear flow boundary layer, due to the existence of critical layers within it~\cite{Brambley12,Dai18}. The effective one-dimensional model and boundary condition \eqref{Bram_BC} discussed in the preceding section presents the advantage to have canonically conserved quantities. To see this, it is valuable to use the Lagrange-Hamilton formalism. This allows us to construct conserved quantities by standard procedures. For instance, the energy is given by the Hamiltonian functional applied on a solution. Our effective model can be obtained from the Lagrangian density  
\be
\mathcal L = \frac12 \left[ (D_t \varphi)^2 - (\p_x \varphi)^2 - 3(\psi - \varphi)^2 + b (D_t \psi)^2 + \delta M_0 (\p_x A) (\p_x^2 \psi) - 2\alpha (\p_t A - \psi) \right] . 
\ee
In this Lagrangian we have introduced an auxiliary field $A(t,x)$ and a Lagrange multiplier $\alpha(t,x)$, which ensures the constraint $\p_t A = \psi$. This procedure allows us to generate a correction with a term in $1/\om$ while having well defined equations in the time domain. Indeed, since the constraint is $\p_t A = \psi$, for a stationary state we will have $A = i \psi/\om$. Requiring vanishing variations of the Lagrangian gives the Euler-Lagrange equations~\footnote{At this level, we considered localized time dependent solutions (finite energy), and hence all boundary terms arising by integration by parts vanish.}, which read 
\bsub \bea
D_t^2 \varphi - \p_x^2 \varphi &=& 3(\psi - \varphi), \\
-D_t b D_t \psi + \frac12 \delta M_0 \p_x^3 A + \alpha &=& 3 (\psi - \varphi), 
\eea \esub
together with the constraint 
\be
\p_t A = \psi, 
\ee
and the equation on the Lagrange multiplier 
\be
\delta M_0 \p_x^3 \psi - 2 \p_t \alpha = 0. 
\ee 
Using the constraint, this last equation integrates into 
\be \label{Lagrange_M}
\alpha = \frac12 \delta M_0 \p_x^3 A. 
\ee
Hence the system becomes 
\bsub \label{1D_Eq_system} \bea
D_t^2 \varphi - \p_x^2 \varphi &=& 3(\psi - \varphi), \\
-D_t b D_t \psi + \delta M_0 \p_x^3 A &=& 3 (\psi - \varphi). 
\eea \esub
We now see that these equations indeed give the effective model obtained by combining \eqref{Bram_BC}, \eqref{Integrated_wave_eq}, and \eqref{Parab_Approx}. 
The Hamiltonian is now obtained by a Legendre transform of the Langrangian. Defining the conjugate momenta 
\bsub \bea
\pi_\varphi &=& D_t \varphi, \\
\pi_\psi &=& b D_t \psi, \\
\pi_A &=& -\alpha, 
\eea \esub
the Hamiltonian density is given by 
\be
\mathcal H = \pi_\varphi \p_t \varphi + \pi_\psi \p_t \psi + \pi_A \p_t A - \mathcal L. 
\ee
This finally gives 
\bea \label{Energy_Integral}
E = \frac12 \int \Big[ (\p_t \varphi)^2 + (1-M_0^2)(\p_x \varphi)^2 &+& b (\p_t \psi)^2 + 3 (\psi - \varphi)^2 - bM_0^2 (\p_x \psi)^2 \nonumber \\
&-& 2\delta M_0 \p_x A \p_x^2 \psi \Big] \d x. 
\eea
In general, it is equally useful to work with a local conservation law of the form~\footnote{Notice that unlike \eq{Energy_Integral}, the local conservation law \eqref{Conservation_law} does not rely on any assumption on boundary terms.}
\be \label{Conservation_law}
\p_t \mathcal E + \p_x J = 0. 
\ee
That can be done with the energy, by defining an energy density $\mathcal E_e$ and current $J_e$ such that $E = \int \mathcal E_e \d x$ and $\p_t \mathcal E_e + \p_x J_e = 0$. However, it is slightly simpler to use the conservation of the symplectic norm. The corresponding density is canonically defined as  
\be \label{norm_density}
\mathcal E_s \doteq -\Im\left(\varphi^* \pi_\varphi + \psi^* \pi_\psi + A^* \pi_A \right) = -\Im\left(\varphi^* D_t \varphi + b \psi^* D_t \psi - A^* \alpha \right). 
\ee
Using the equation of motion \eqref{1D_Eq_system}, we directly show that the corresponding current is given by 
\be \label{Current_eq}
J_s = \Im \left[ \varphi^* \p_x \varphi - M_0 \varphi^* D_t \varphi - M_0 b \psi^* D_t \psi + \frac12 \delta M_0 \left(\p_t A^* \p_x^2 A - A^* \p_x^2\p_t A - \p_t \p_x A^* \p_x A\right) \right]. 
\ee
The local conservation law $\p_t \mathcal E_s + \p_x J_s = 0$ implies that any localized solution have a conserved total norm $\int \mathcal E_s \d x$. Although closely related, this norm is different from the energy. In fact, it is a generalization of the concept of \emph{wave action} defined as the ratio of the energy over the frequency $E/\om$ (see~\cite{Vanneste99,Buhler} for a general discussion and~\cite{Coutant13,Coutant16b} for its use in a similar context). As one can directly verify, for a monochromatic wave, $E = \om \int \mathcal E_s \d x$. The main difference is that the wave action is defined in a WKB limit, in which case it becomes conserved (it is an adiabatic invariant), while the total norm is always conserved, without any approximation. 

We now evaluate the current $J_s$ and the density $\mathcal E_s$ for a single frequency wave $e^{-i \om t + i k x}$, assuming $b$ constant. Using the dispersion relation \eqref{1D_Disp_rel} and the equations of motion \eqref{1D_Eq_system}, we obtain 
\be \label{Norm_density_PW}
\mathcal E_s(\om, k_j) = (\om - M_0 k_j) + \frac19 \left( b (\om - M_0 k_j) - \frac{\delta M_0 k_j^3}{2\om^2}\right) \left(-(\om - M_0 k_j)^2 + k_j^2 + 3 \right)^2, 
\ee
where $j \in \{\s, \+, \u, \v, \- \}$. A direct evaluation of \eqref{Norm_density_PW} shows that $k_\-$ is of negative norm, while the four other roots have a positive norm (and the same follows for the energy since $E = \om \int \mathcal E_s \d x$). To see this more simply, one can notice that in the correction term in $\delta$, one can replace $-M_0 k_j^3$ by $(\om-M_0 k_j)^3/M_0^2$ to a good approximation (i.e. in the same limit the boundary condition \eqref{Bram_BC} was obtained -- see App.~\ref{BC_App}). Doing so, we see that the sign of $\mathcal E_s$ is that of the intrinsic frequency $\om - M_0 k_j$, which is positive for all roots but $k_\-$ (we present in App.~\ref{Roots_App} asymptotic expressions for the norm and energy). The second important point to notice is that for a single frequency wave, the conservation law \eqref{Conservation_law} takes the form  
\be \label{Single_Freq_Conservation}
J_s = v_g \mathcal E_s, 
\ee
where $v_g = \p_k \om$ is the group velocity of the corresponding mode. To derive this identity, the idea is to apply the conservation law \eqref{Conservation_law} to a tight wave packet $\varphi = f(t,x) e^{-i \om t + i k x}$, where $f$ is a slowly varying envelope, which depends essentially on the variable $t - x/v_g$. Since $J_s$ and $\mathcal E_s$ depend on $t$ and $x$ only through $f$, the above identity \eqref{Single_Freq_Conservation} follows.
%~\footnote{For a more detailed proof of this identity, we refer for instance to App.~B of~\cite{Coutant18c}. Although this was derived in a very different context, the proof applies immediately to the present case.}. 
This equality shows that for a negative energy wave ($\mathcal E_s < 0$), the wave propagate in the opposite direction as the energy current.

%%%%%%%%%%%%%%%%%%%%%%%%%%%%%%%%%%%%%%%%%%%%%%%%%%%
%LASER EFFECT IN 1D MODEL
%%%%%%%%%%%%%%%%%%%%%%%%%%%%%%%%%%%%%%%%%%%%%%%%%%%
\section{Laser effect}
\begin{figure}[!ht]
\centering
\framebox{\includegraphics[width=0.75\columnwidth,trim=0 0 0 0,clip]{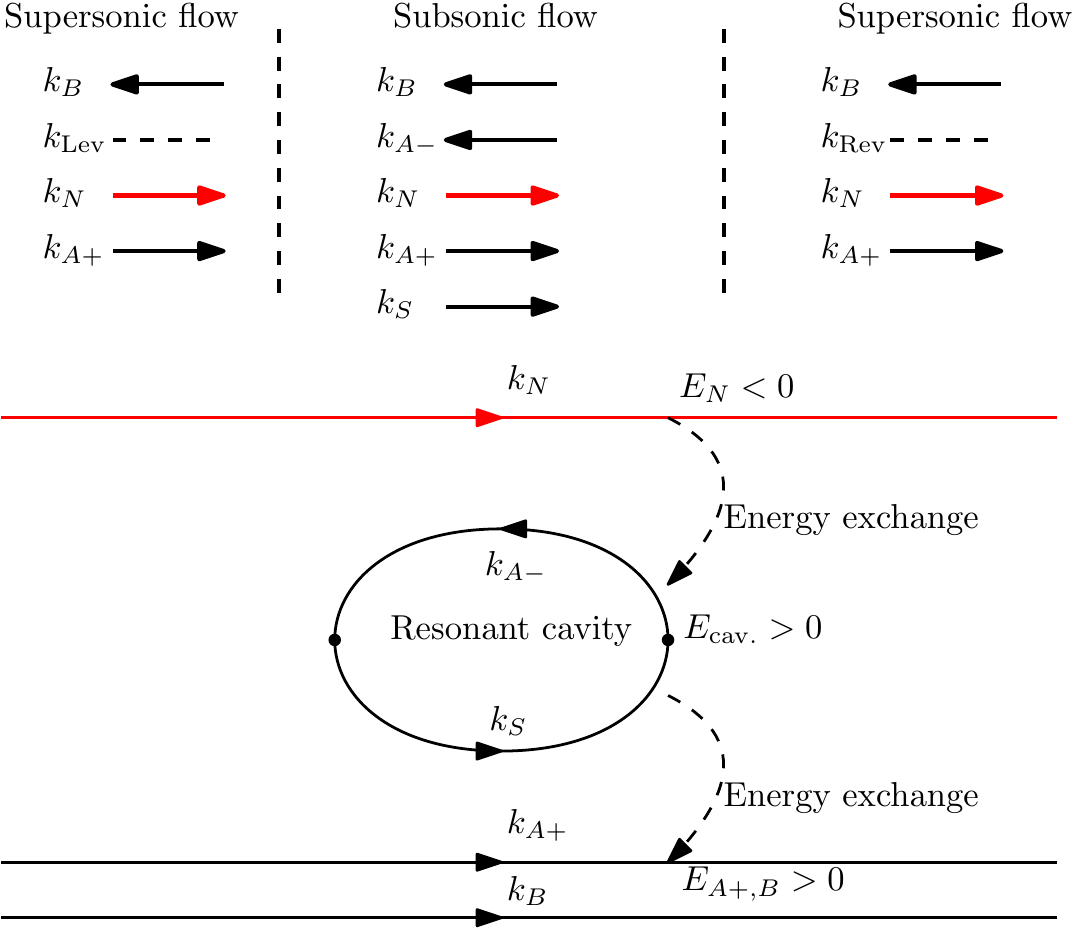}}
\caption{Schematic representation of the laser mechanism. Black arrows and lines carry a positive energy while red carry a negative energy. The mechanism keeps the total energy $E_N+E_{\rm cav.}+E_{A+}+E_B$ constant. 
}
\label{BHL_Fig} 
\end{figure}

We now consider a configuration with two transsonic transitions. The flow is supersonic, except in a region of size $L$ (in the units of transverse height) where it is subsonic. In this case, modes can be trapped in the central region, triggering the black hole laser instability. As illustrated in Fig.~\ref{BHL_Fig}, the core of the mechanism is the coupling between a resonant cavity ($0 < x < L$) and the negative energy wave. When the cavity modes radiate energy through the negative energy wave, it results in an increase of energy in the cavity. In general there is a competition between the coupling to the negative energy wave, which tends to increase the cavity energy, and the coupling to the acoustic mode $k_{A+}$, which takes energy away. On a transsonic transition, the coupling to $k_{A+}$ is generally quite smaller than to $k_N$ (see e.g. Fig.~7 in~\cite{Auregan15b}) and hence the cavity is unstable. When taking into account a finite boundary layer, the presence of an extra mode provides an additional source of damping. As we shall see in section~\ref{Bram_Sec}, the presence of the boundary layer generally reduces the laser instability. 

\begin{figure}[!ht]
\centering
\includegraphics[width=0.6\columnwidth]{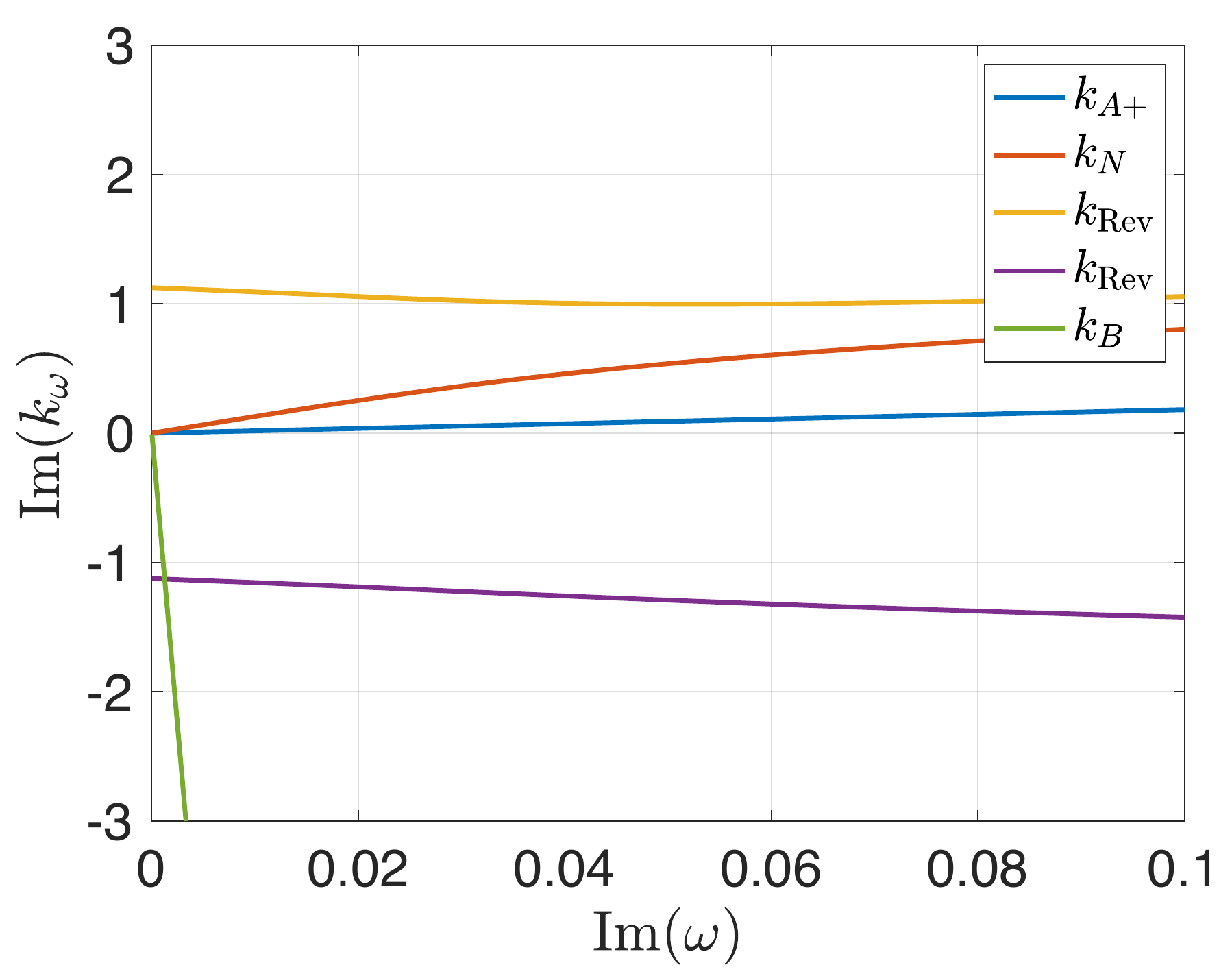}
\caption{Evolution of $\Im(k_\om)$ when $\Im(\om)$ changes from large positive values to 0 for the modes in a supersonic region. We have taken $M_0=0.3$, $b = 15$, $\delta=0.005$, and $\Re(\om)=0.03$. Using the Briggs-Bers criterion, the sign of $\Im(k_\om)$ gives the direction of propagation of the corresponding mode. (We have verified that the signs of $\Im(k_\om)$ don't change when increasing $\Im(\om)$ further.)
}
\label{BriggsBers_Fig} 
\end{figure}

This instability is encoded in a set of complex eigenfrequencies of positive imaginary parts~\cite{Coutant10}, hence leading to a growing behavior. To find them, we look for stationary solutions of the form 
\bsub \bea \label{Statio_Modes}
\varphi(t,x) &=& \varphi_\om(x) e^{-i \om t}, \\
\psi(t,x) &=& \psi_\om(x) e^{-i \om t}, 
\eea \esub
which obey the stationary equations 
\bsub \label{1D_Mode_Eq} \bea
-(\om + i M_0 \p_x)^2 \varphi - \p_x^2 \varphi &=& 3(\psi - \varphi), \\
(\om + i M_0 \p_x) b (\om + i M_0 \p_x) \psi + \frac{i\delta M_0}{\om} \p_x^3 \psi &=& 3 (\psi - \varphi). 
\eea \esub
We now look for $\om \in \mathbb C$ such that there exist a purely outgoing solution. To properly define this, we proceed in the following way. When $\Im(\om) > 0$, we select solutions such that $\varphi$ and $\psi$ are both decaying for $x \to \pm \infty$. In other words, unstable modes are spatially localized. We then obtain the condition for $\Im(\om) < 0$ by analytic continuation in the lower half plane. This procedure, identical to that of~\cite{Coutant10}, is equivalent to the Briggs-Bers prescription~\cite{Briggs64,Bers83,Crighton91}. Indeed, we verify that when starting with $\Im(\om) \to +\infty$, the sign of $\Im(k_\om)$ do not change while lowering $\Im(\om)$. This is shown in Fig.~\ref{BriggsBers_Fig}. Moreover, when $\Im(\om) \to 0$, the identity
\be
k(\om_r+i \epsilon) = k(\om_r) + i\epsilon/v_g + O(\epsilon^2) 
\ee
allows us to show that the sign of $\Im(k_\om)$ is the same as the group velocity. In addition, the spectrum is invariant under the discrete symmetry 
\be \label{Sp_Sym}
\om \to -\om^*. 
\ee
This comes from the fact that the boundary conditions we consider are unchanged under complex conjugation. Then, taking the complex conjugate of the mode equation \eqref{1D_Mode_Eq} gives a one-to-one mapping between a solution for $\om$ and $-\om^*$. This symmetry naturally divides the spectrum into two classes: purely imaginary frequencies, which are maximally symmetric and encode a static instability, and complex frequencies with non-zero real part, which come in pairs $(\om,-\om^*)$ and correspond to a dynamical instability~\footnote{This distinction is similar to the spectrum of $\mathcal P \mathcal T$ symmetric Hamiltonian, which are purely real only when they are maximally symmetric~\cite{Bender05}. In fact, the symmetry \eqref{Sp_Sym} comes from the symmetry $(\om,k) \to (-\om,-k)$ of the dispersion relation \eqref{1D_Disp_rel}, which can be seen as a local version of $\mathcal P \mathcal T$ symmetry.}. Keeping that symmetry in mind, we will in the following focus on the part of the spectrum $\Re(\om) > 0$. 

We now analyse the set of complex eigen-frequencies for two abrupt changes in the wall compliance separated by a distance $L$. For this we assume that the length of the tubes vary along the wall as 
\be
b(x) = \left\{ \bal & b_O \quad (x<0), \\
& b_C \quad (0< x < L), \\
& b_O \quad (L < x). 
\eal \right.
\ee
The values are chosen such that the flow is supersonic on the outer region ($M_0 \sqrt{1+b_O} > 1$) but subsonic in the central region ($M_0\sqrt{1+b_{C}} < 1$). We consider separately the case of a vanishing boundary layer (Ingard-Myers condition) and a nonzero one (boundary condition \eqref{Bram_BC}).

%%%%%%%%%%%%%%%%%%%%%%%%%%%%%%%%%%%%%%%%%%%%%%%%%%%
%INGARD MYERS
%%%%%%%%%%%%%%%%%%%%%%%%%%%%%%%%%%%%%%%%%%%%%%%%%%%

\subsection{Complex spectrum for a vanishing boundary layer}
\begin{figure}[!ht]
\centering
\includegraphics[width=0.49\columnwidth]{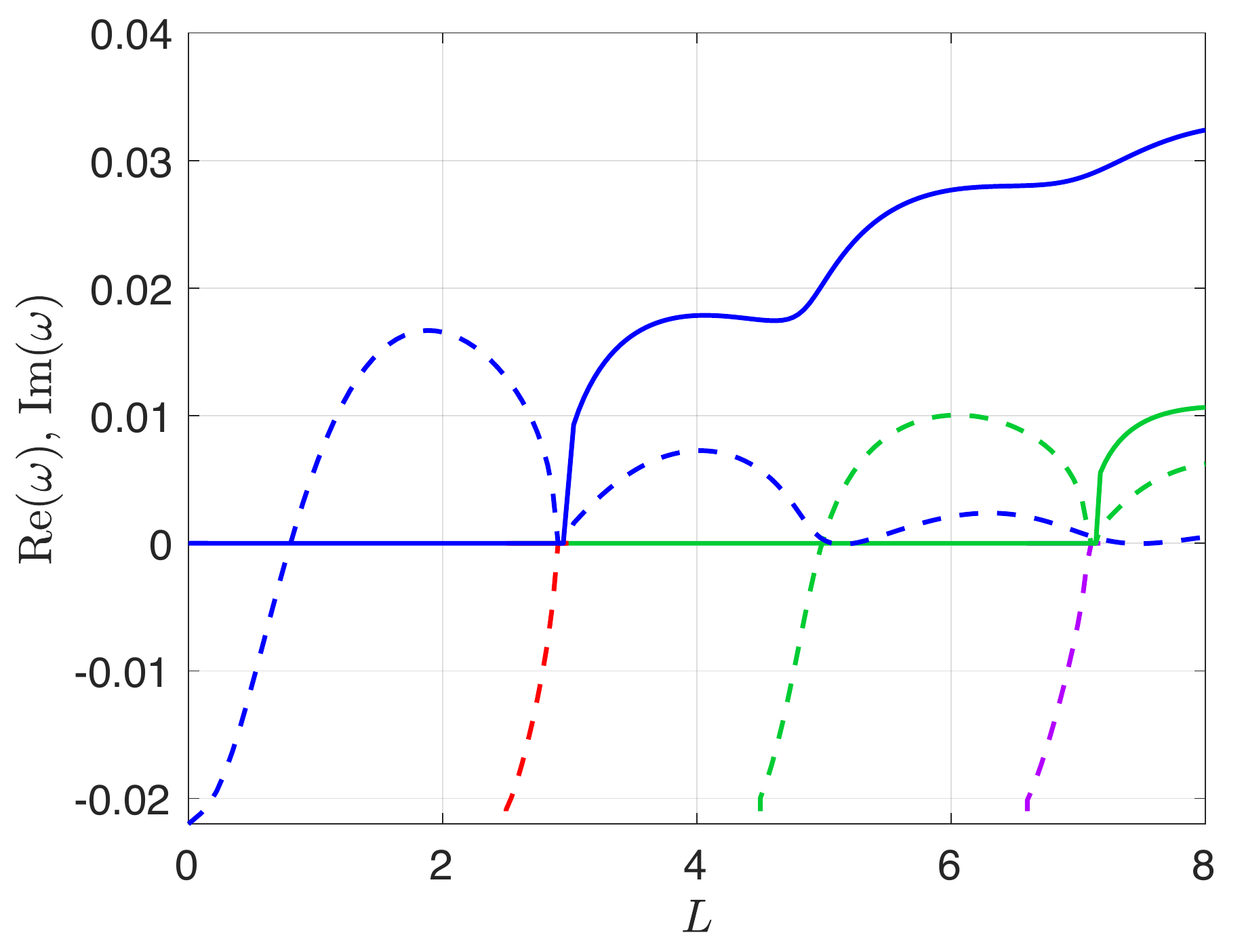}
\includegraphics[width=0.49\columnwidth]{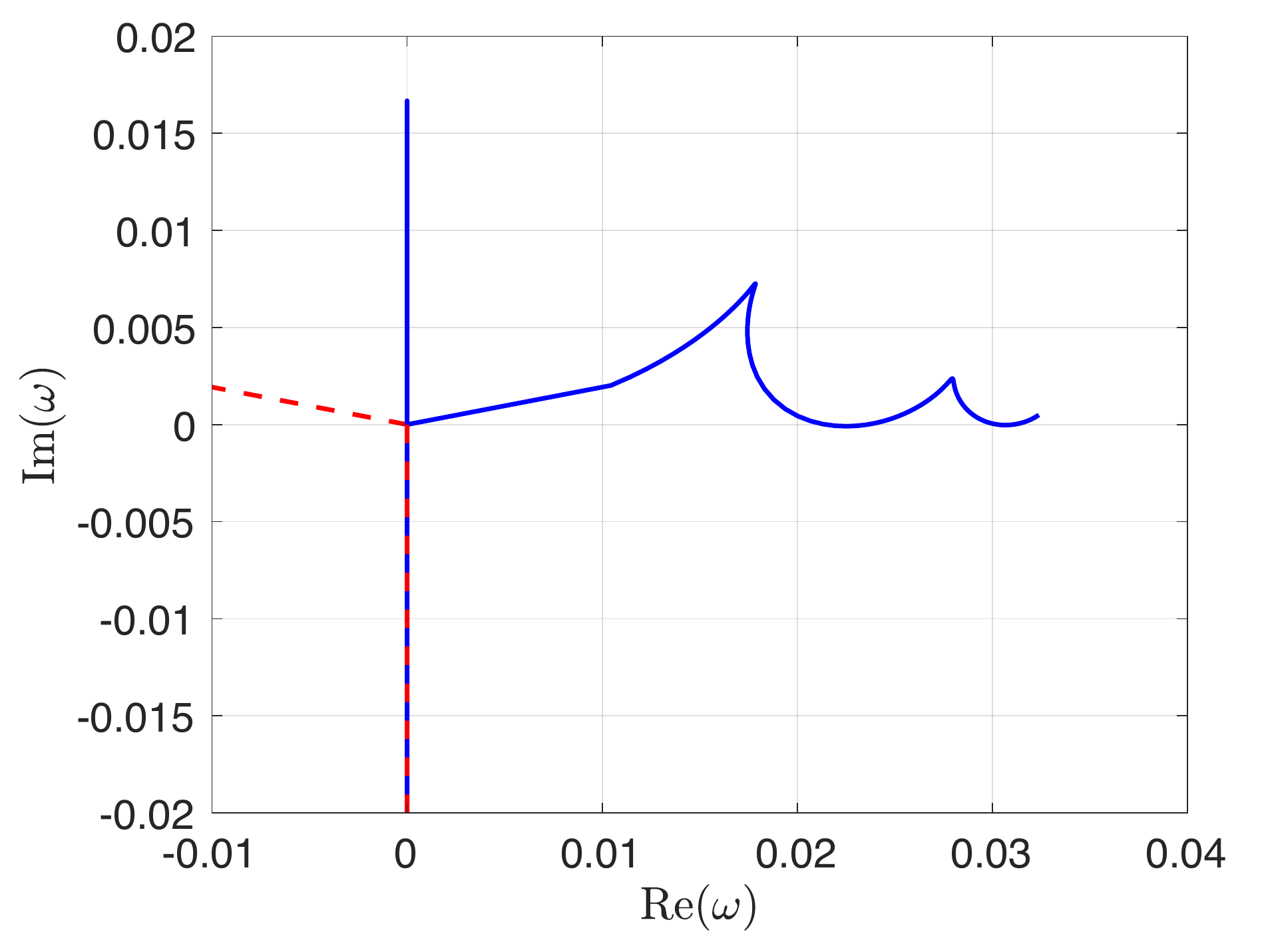}
\caption{Trajectory of unstable modes when $L$ varies from $0$ to $8$ with $M_0=0.3$, $b_C = 6$, $b_O = 15$. On the left: solid (resp. dashed) lines show the evolution of the real (resp. imaginary) parts of $\om$ as a function of $L$. On the right: trajectories in the complex plane of the first pair of modes. 
}
\label{BHL_Cplane_Fig} 
\end{figure}

\begin{figure}[!ht]
\centering
\includegraphics[width=0.49\columnwidth]{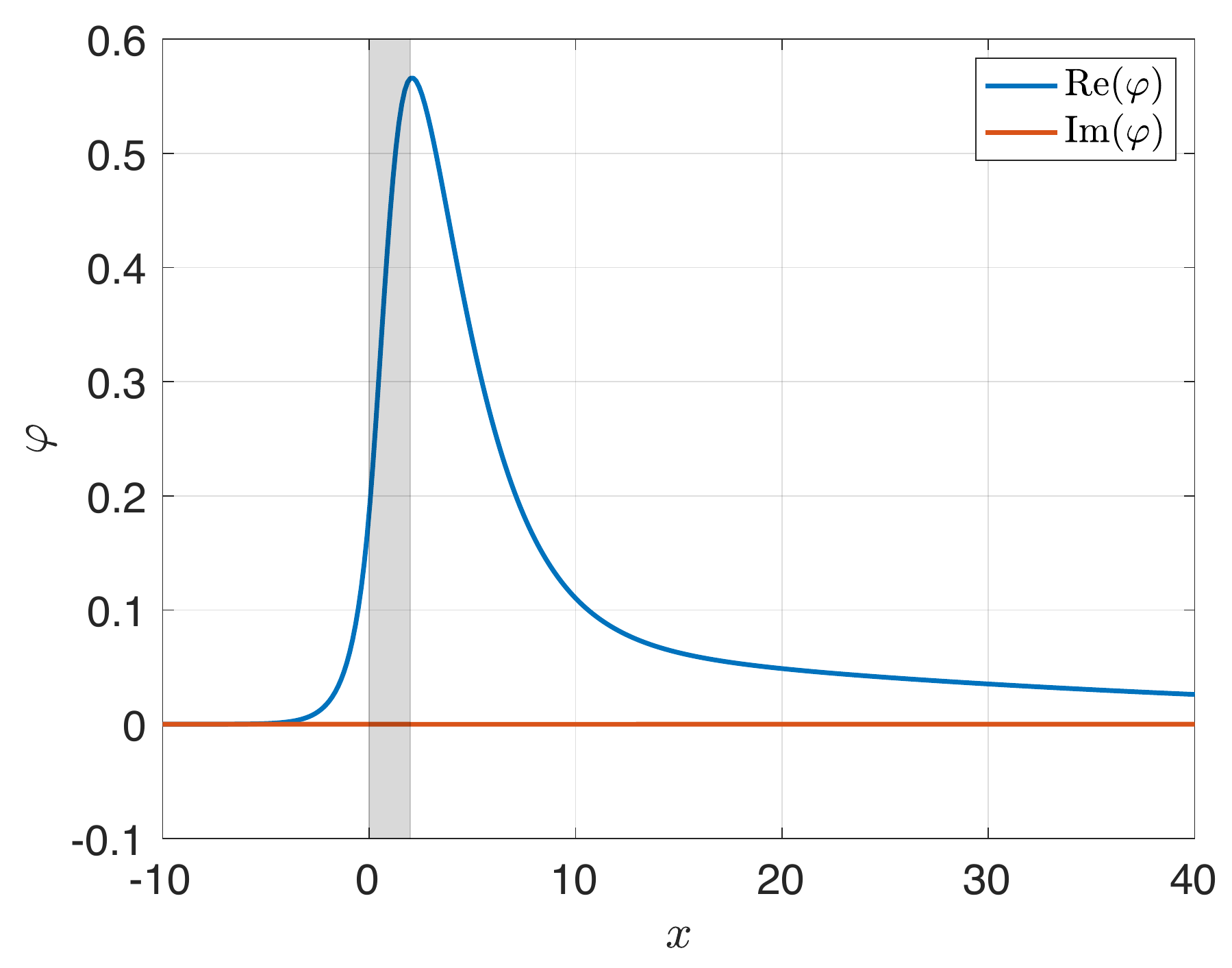}
\includegraphics[width=0.49\columnwidth]{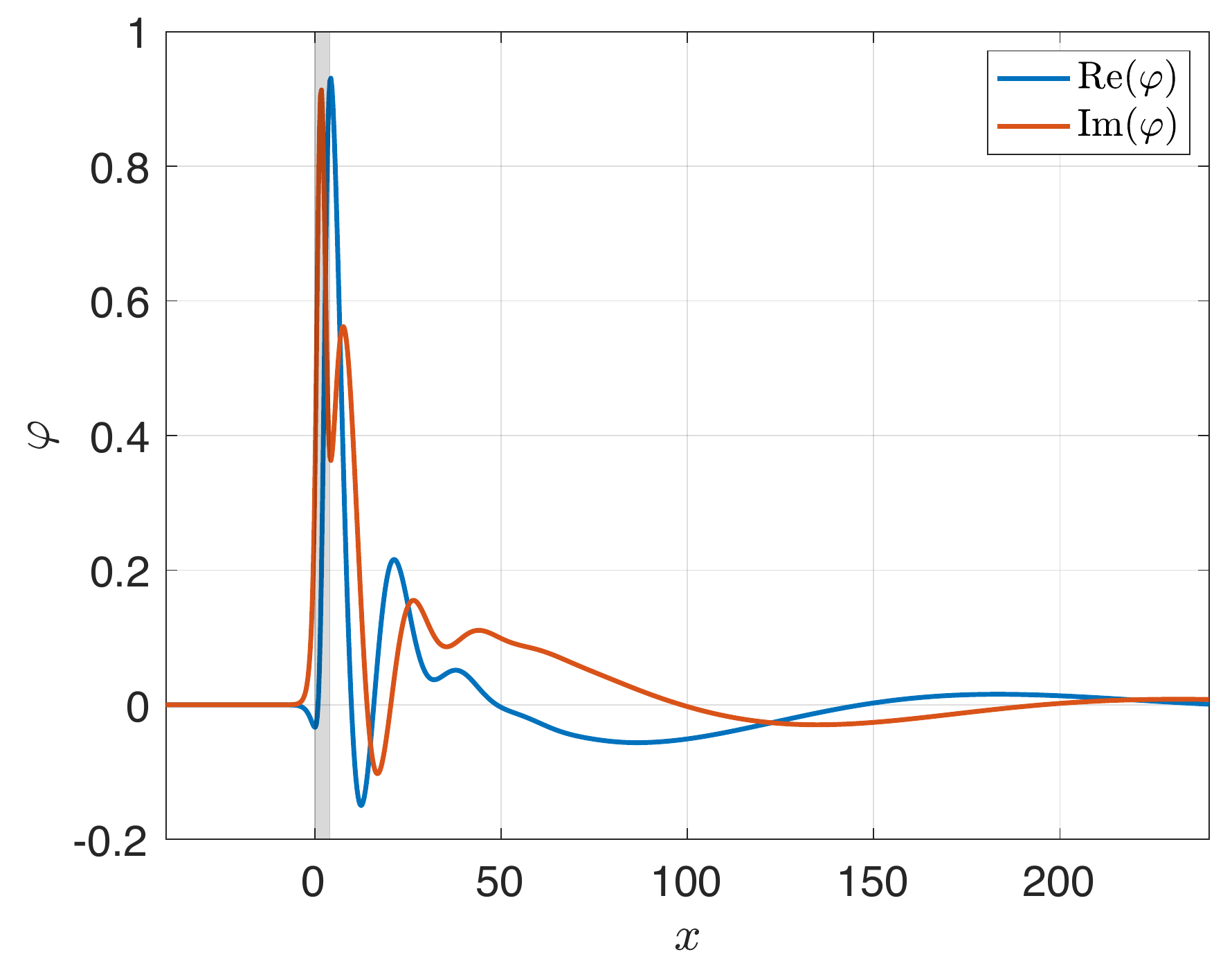}
\caption{Profile of the first unstable mode (blue curved in Fig.~\ref{BHL_Cplane_Fig}) with $M_0=0.3$, $b_C = 6$, $b_O = 15$. On the left: $L=2$, the mode is a static instability. On the right: $L=4$, the mode is a dynamical instability. The grey shading delimitates the subsonic region. 
}
\label{BHL_ModesIMbc_Fig} 
\end{figure}

In each region, the solutions of the mode equation \eqref{1D_Mode_Eq} are superpositions of exponentials $e^{ikx}$ where $k$ satisfies the dispersion relation \eqref{1D_Disp_rel}. To satisfy outgoing boundary conditions, we look for solutions of the form 
\be \label{Complex_Eigen_Modes}
\varphi_\om(x) = \left\{ \bal & a_L \frac{e^{i k_{\rm Lev} x}}{\sqrt{J(\om,k_{\rm Lev})}} \quad (x<0), \\
& a_\- \frac{e^{i k_\- x}}{\sqrt{J(\om,k_\-)}}  + a_\v \frac{e^{i k_\v x}}{\sqrt{J(\om,k_\v)}} + a_R \frac{e^{i k_{\rm Rev} x}}{\sqrt{J(\om,k_{\rm Rev})}} \quad (L<x), 
\eal \right.
\ee
where $k_{\rm Lev}$ (resp. $k_{\rm Rev}$) is the evanescent mode on the left (resp. right). At both interfaces, $x=0$ and $x=L$, we must use proper continuity conditions. Because the boundary layer correction (see \eq{Bram_BC}) changes the order of the equation, going from 4 to 5, one must carefully discuss the Ingard-Myers condition and Brambley condition separately. For $\delta = 0$, the mode equations \eqref{1D_Mode_Eq} imply that the fields $\varphi$, $\p_x \varphi$, $\psi$, and $b (\om + i M_0 \p_x) \psi$ are continuous. Notice that the last condition ensures that the current is conserved across the interfaces. These conditions give a linear relation between the coefficients $a_j$ in equation \eqref{Complex_Eigen_Modes}. We numerically evaluate the determinant of the corresponding system, and look for its zeros. We display the results for the spectrum on Fig.~\ref{BHL_Cplane_Fig}, and the eigen-modes on Fig.~\ref{BHL_ModesIMbc_Fig}. 

When the size $L$ of the subsonic region increases, more and more unstable modes appear. Each new mode appear by following two steps. It first comes out of the origin ($\om=0$) to be purely imaginary: it is a static instability. It then comes back to the origin and is converted into a complex frequency (both real and imaginary parts are non-zero), hence becoming a dynamical instability. This laser instability has several peculiar properties which we underline now: 
\bi
\item The studied configuration is not always unstable. If the trapping region is `too small,' no unstable mode exist. For the parameters of Fig.~\ref{BHL_Cplane_Fig}, we see that this is the case for $L \lesssim 0.8$. Moreover, it is noticeable that the purely imaginary resonance that precedes the static instability for small $L$ tends to a finite value in the limit $L\to 0$ (this is due to the presence of a branch point at this value of $\om \in \mathbb C$). 
\item There are generically both static and dynamical unstable modes. However, the dominating one can be either one, depending on the parameters of the duct configuration. This is due to interference effects, which generate an oscillating behavior of the imaginary part of the complex frequency when varying external parameters (see section IV.~C in~\cite{Coutant10}). Notice that these interferences would be reduced if the impedance values are different on both supersonic regions. 
\item When $\Re(\om) > \om_\M$ we no longer see any unstable mode. This \emph{a priori} nontrivial, since negative energy waves still exist pass this threshold frequency. The reason is that when $\om > \om_\M$, the negative energy waves can transfer energy only to the mode $k_\v$, but the latter is essentially decoupled from the other ones. This means that the transmission is almost perfect at both interfaces, which prevents the unstable mechanism. 
\ei
As we mentioned in the introduction, a similar laser instability (`the black hole laser') has been studied in Bose-Einstein condensates, in theory~\cite{Coutant10,Finazzi10,Michel13,Finazzi14} but also observed in an experiment~\cite{Steinhauer14}. The above properties of the acoustic black hole laser are very close to the one identified in Bose-Einstein condensates. This is somewhat surprising, since in the latter, it is the negative energy wave that is trapped in the cavity (compare with Fig.~\ref{BHL_Fig}). This is due to a different nature of the dispersion relation in both media: in condensates, acoustic waves propagate faster when the wavelength decreases, while they propagate slower in acoustic liners (see Fig.~\ref{1DeffDisp_Fig}). However, that the two configurations share many similarities was anticipated in~\cite{Coutant11} using a semiclassical symmetry argument. 

\begin{figure}[!ht]
\centering
\includegraphics[width=0.49\columnwidth]{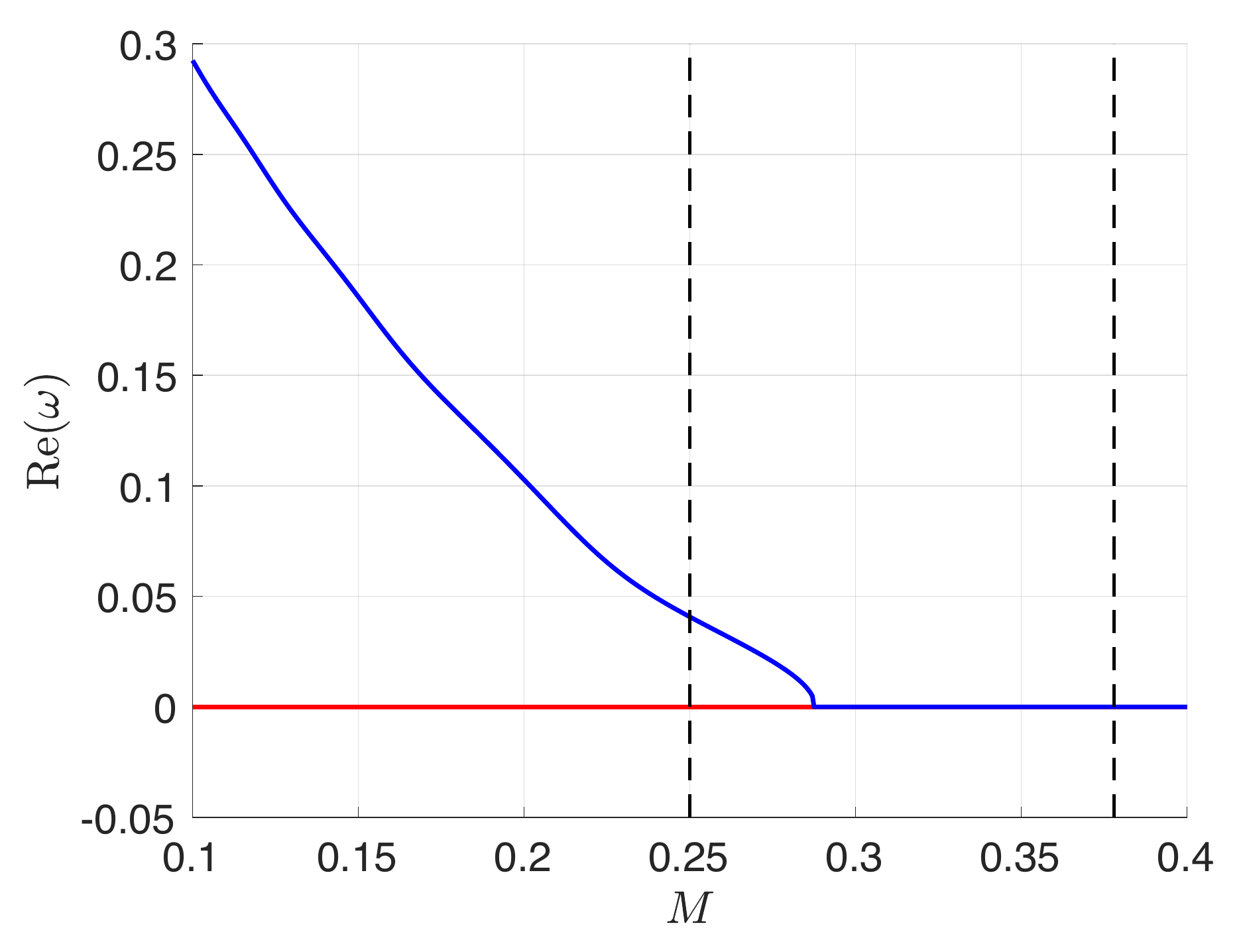}
\includegraphics[width=0.49\columnwidth]{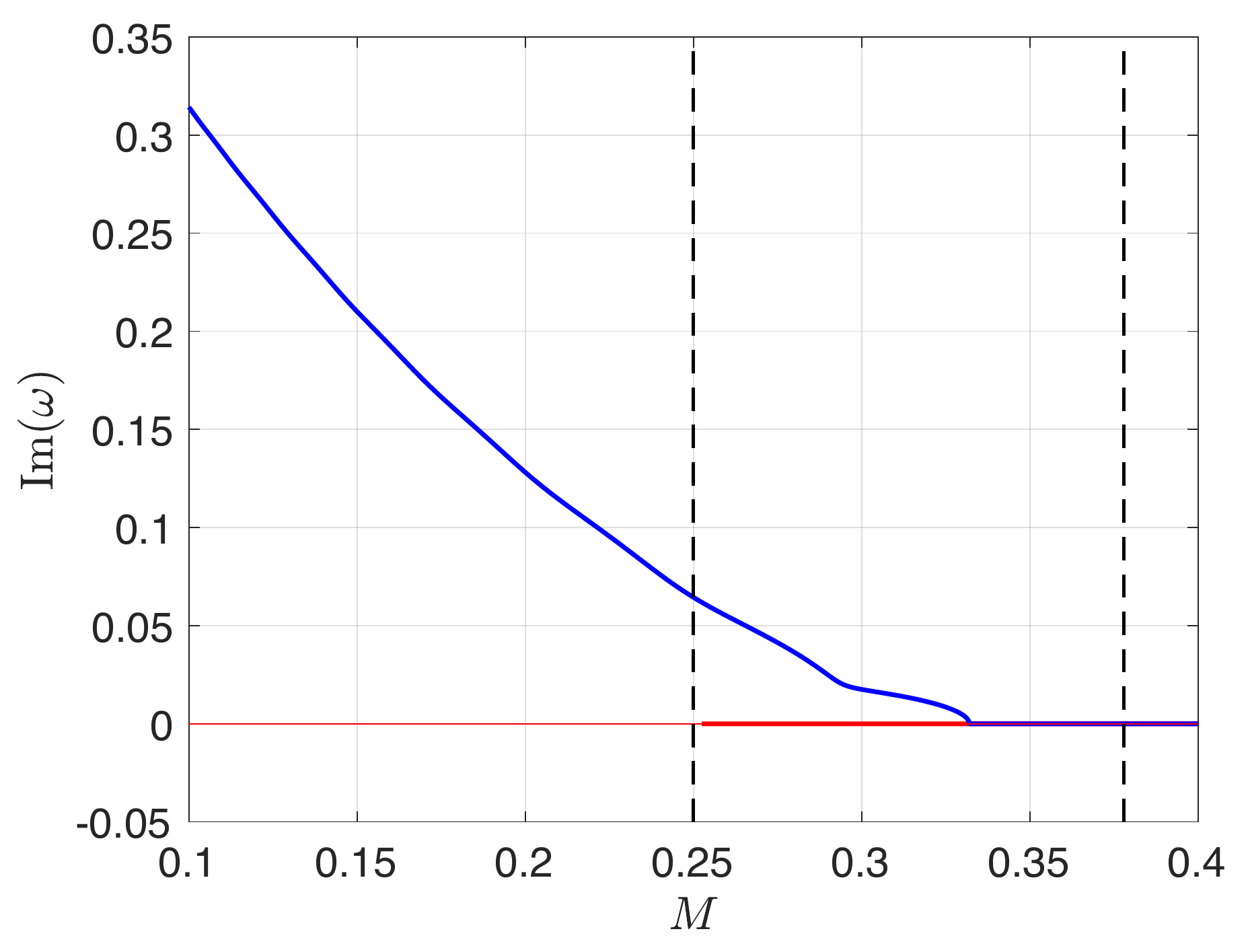}

\includegraphics[width=0.49\columnwidth]{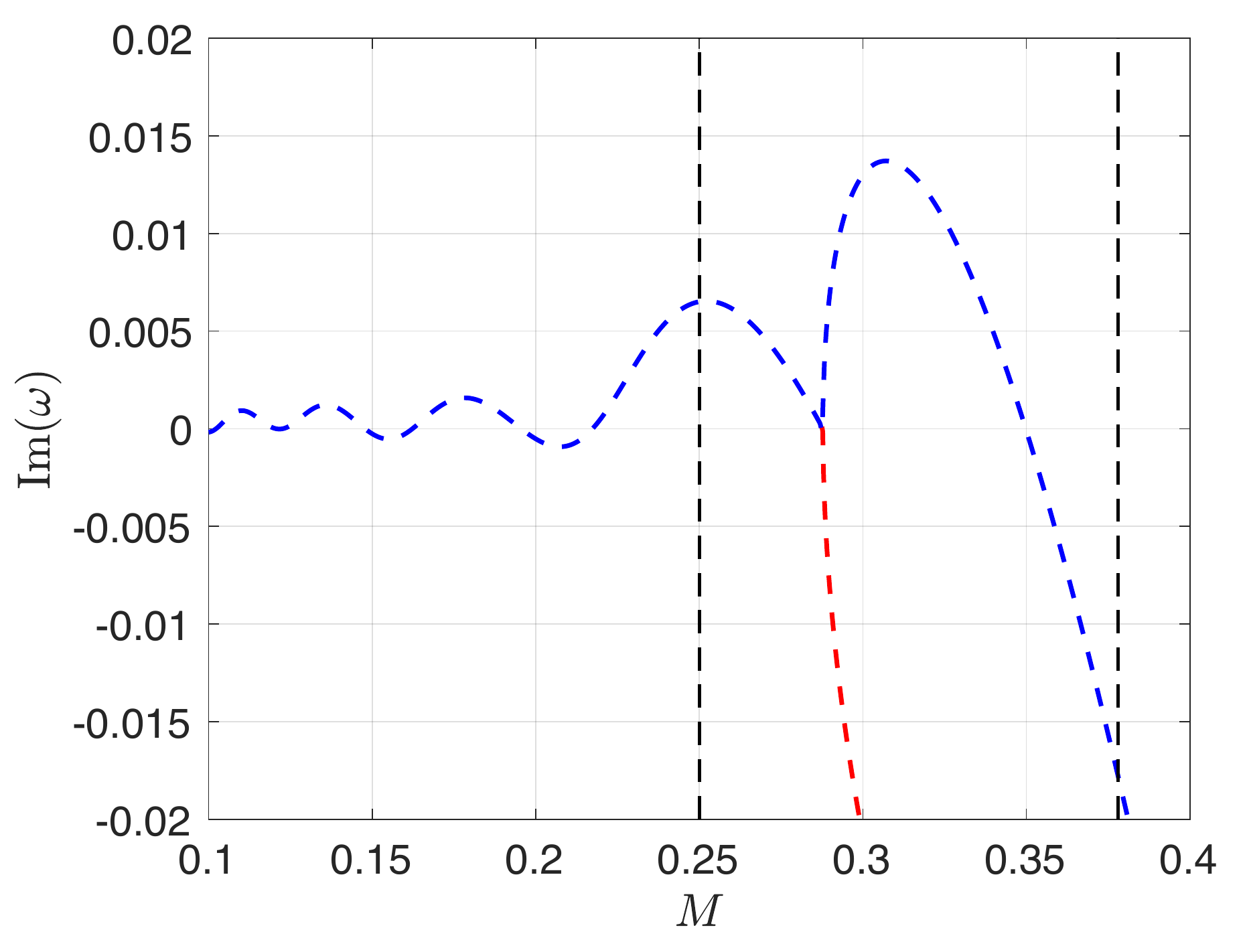}
\includegraphics[width=0.49\columnwidth]{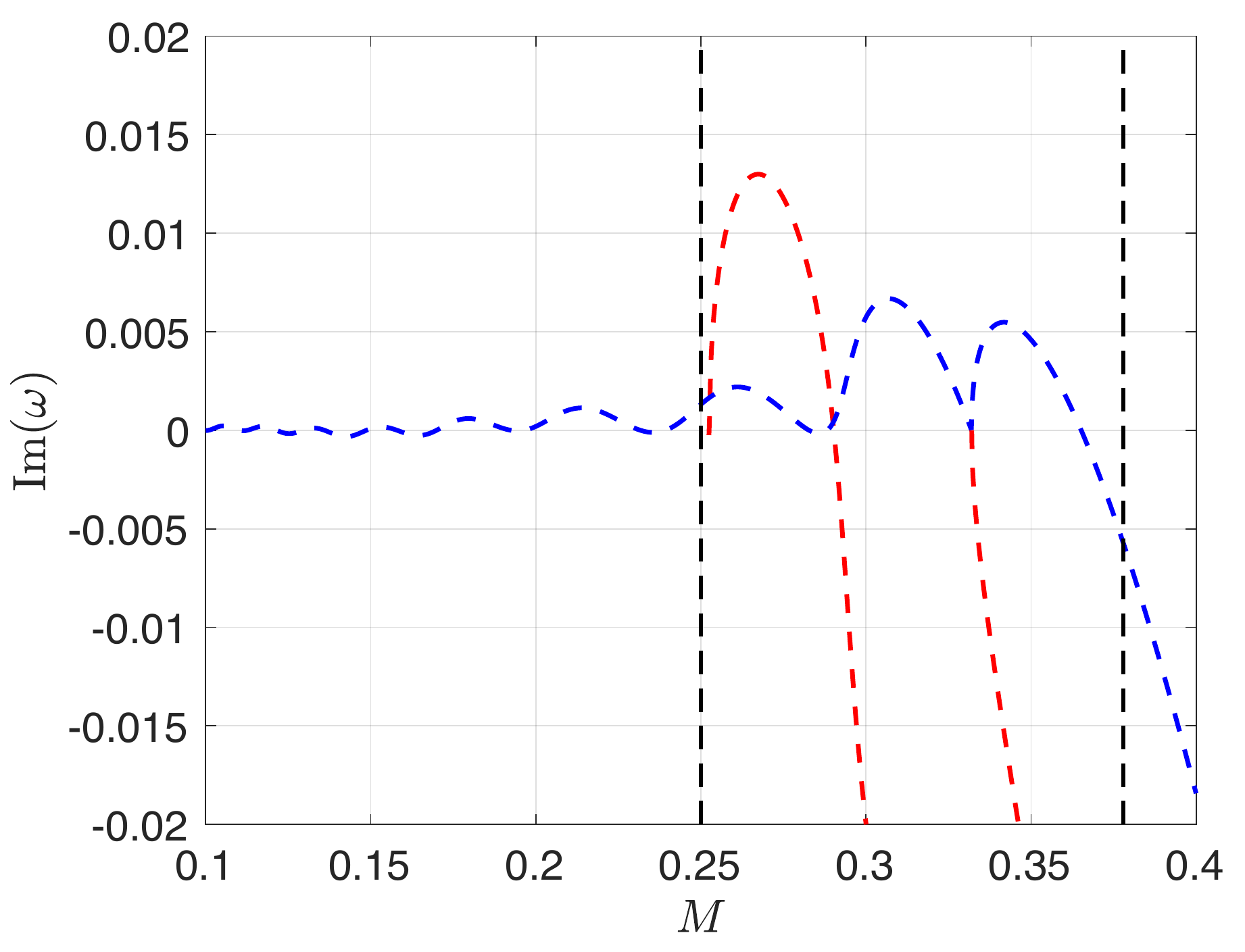}
\caption{Evolution of unstable modes when varying $M_0$. We have chosen $b_C = 6$, $b_O = 15$. On the left: $L=2.5$, on the right: $L=4.5$. Solid lines are real parts, and dashed lines are imaginary parts. The black dashed lines shows the values of $M_0$ below which the flow is everywhere subsonic ($M_0<0.25$) and above which it is everywhere supersonic ($M_0>0.38$). 
}
\label{BHL_Mplot_Fig} 
\end{figure}

We now briefly discuss the influence of the Mach number on the laser effect. In Fig.~\ref{BHL_Mplot_Fig}, we have shown the evolution of unstable modes when varying $M_0$ for two values of $L$. First, we recall that if $M_0 < 1/\sqrt{1+b_O}$, the flow is everywhere subsonic. On the contrary, when $M_0 > 1/\sqrt{1+b_C}$, the flow is everywhere supersonic. The instability can \emph{a priori} exist in these regimes, since both negative and positive energy waves are present. However, it is in general much less strong. This is because transsonic flows couples these waves together much more efficiently. For instance, in Fig.~\ref{BHL_Mplot_Fig} we see that everywhere subsonic flows alternate between being stable and unstable, but the growth rate is significantly smaller that in the doubly transsonic case. In the latter regime, we observe that at low Mach, the instability is dynamical, and it becomes static at higher values of $M_0$. 

%%%%%%%%%%%%%%%%%%%%%%%%%%%%%%%%%%%%%%%%%%%%%%%%%%%
%BRAMBLEY CONDITION
%%%%%%%%%%%%%%%%%%%%%%%%%%%%%%%%%%%%%%%%%%%%%%%%%%%
\subsection{Complex spectrum for a finite boundary layer}
\label{Bram_Sec}
We compute the spectrum again, but using a finite boundary layer. Since there is an extra solution $k_\s$ of the dispersion relation \eqref{1D_Disp_rel}, purely outgoing modes now read  
\be \label{Complex_Eigen_Modes_Bbc}
\varphi_\om(x) = \left\{ \bal & a_L \frac{e^{i k_{\rm Lev} x}}{\sqrt{J(\om,k_{\rm Lev})}} + a_\s \frac{e^{i k_\s x}}{\sqrt{J(\om,k_\s)}} \quad (x<0), \\
& a_\- \frac{e^{i k_\- x}}{\sqrt{J(\om,k_\-)}}  + a_\v \frac{e^{i k_\v x}}{\sqrt{J(\om,k_\v)}} + a_R \frac{e^{i k_{\rm Rev} x}}{\sqrt{J(\om,k_{\rm Rev})}} \quad (L<x).
\eal \right.
\ee
Since the order of the equation has changed, so did the continuity conditions at the interfaces. Inspection of the mode equation \eqref{1D_Mode_Eq} shows that $\varphi$, $\p_x \varphi$, $\psi$, and $\p_x \psi$, and $b(\om + iM_0 \p_x)\psi + \frac\delta\om \p_x^2 \psi$ are continuous. Again, the last condition corresponds to the continuity of the current \eqref{Current_eq}. 

\begin{figure}[!ht]
\centering
\includegraphics[width=0.8\columnwidth]{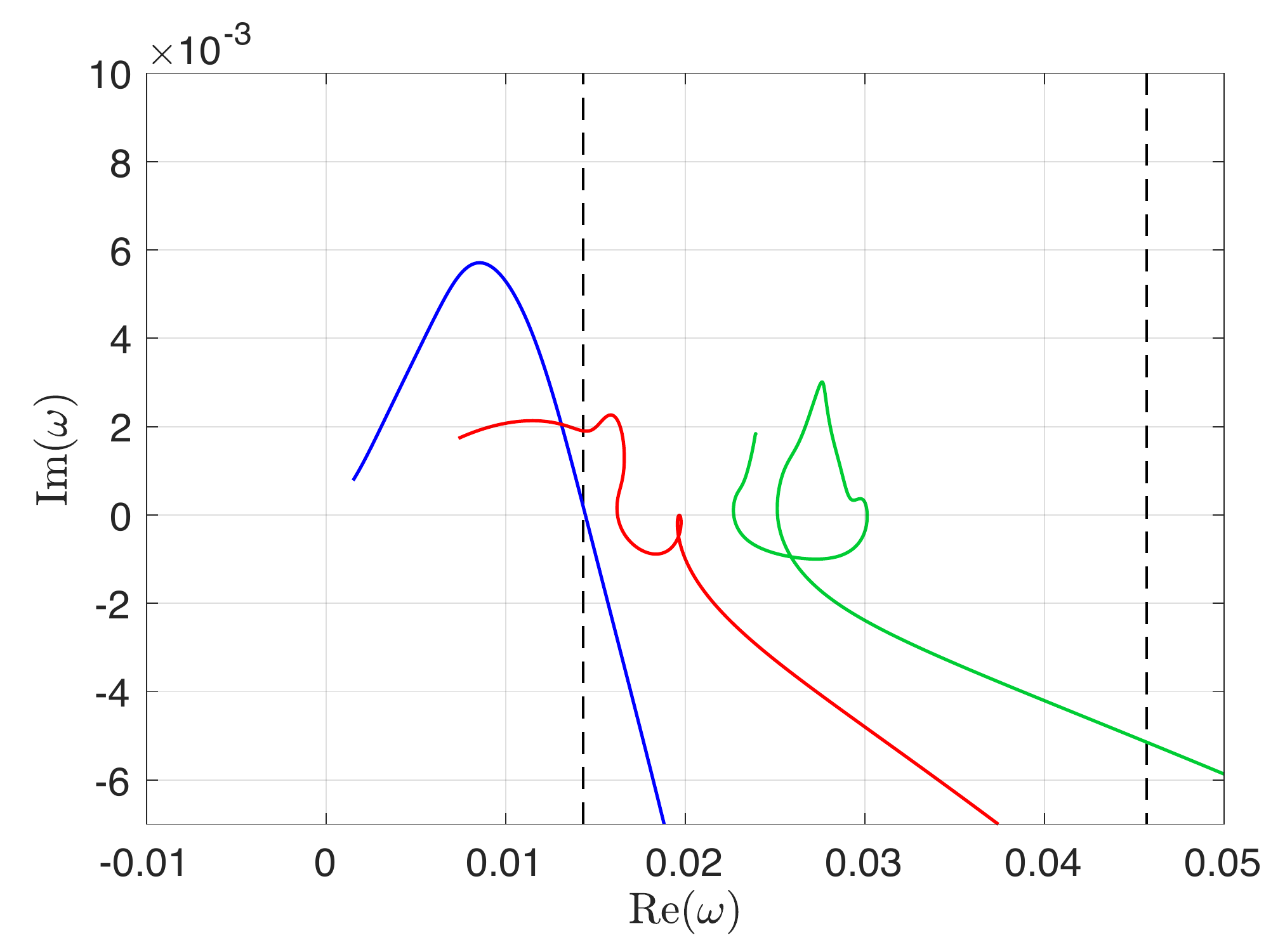}
\caption{Trajectory of unstable modes when $L$ varies from $0$ to $8$ with $M_0=0.3$, $b_C = 6$, $b_O = 15$, and $\delta=0.005$. We have shown the first, third and fifth mode (the second and fourth are stable or have a much smaller imaginary part, hence are subdominant for the laser instability). The dashed black lines mark $\Re(\om) = \om_\m$ and $\Re(\om) = \om_\M$. 
}
\label{BHL_Cplane_Bbc_Fig} 
\end{figure}

\begin{figure}[!ht]
\centering
\includegraphics[width=0.49\columnwidth]{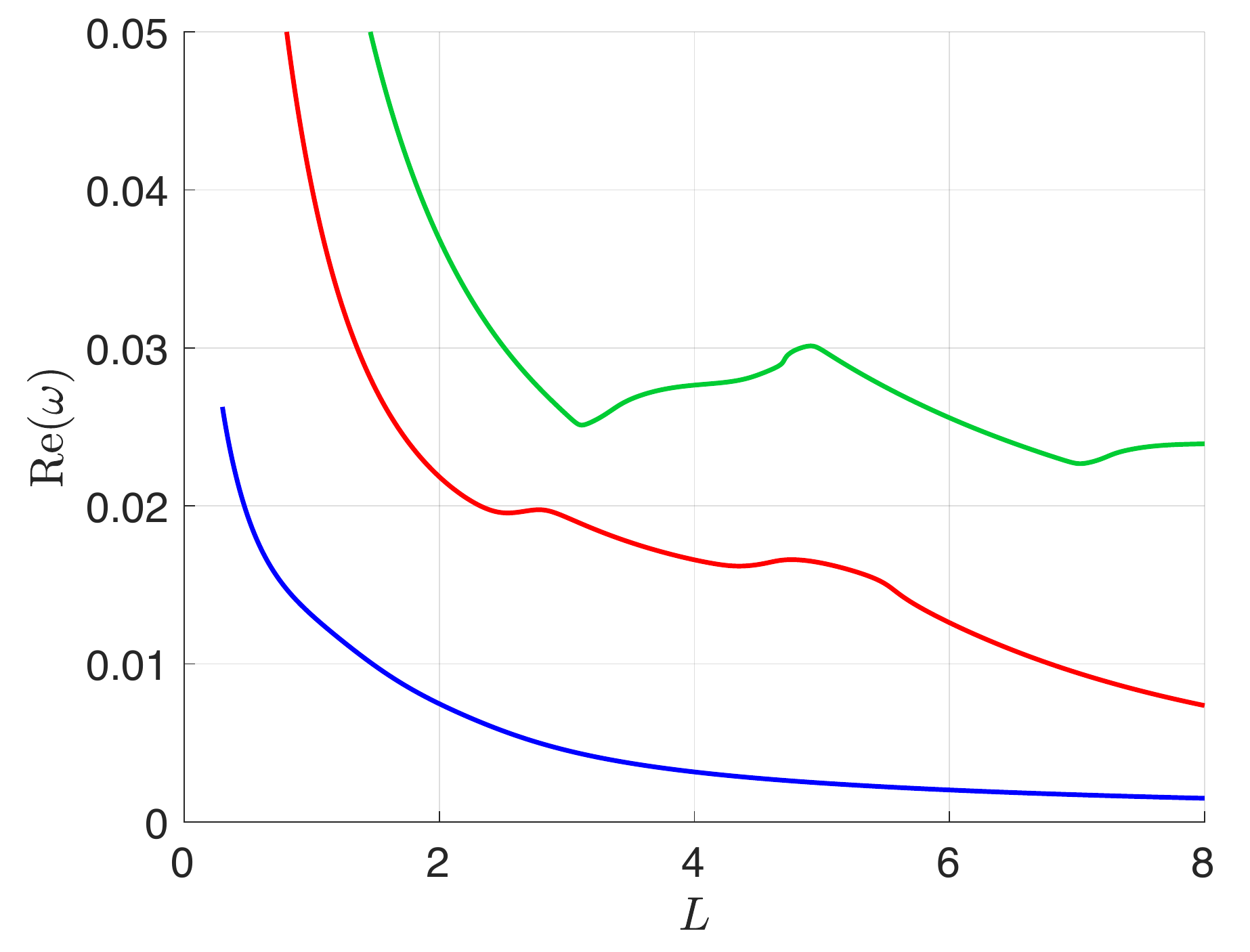}
\includegraphics[width=0.49\columnwidth,trim=0 8pt 0 0]{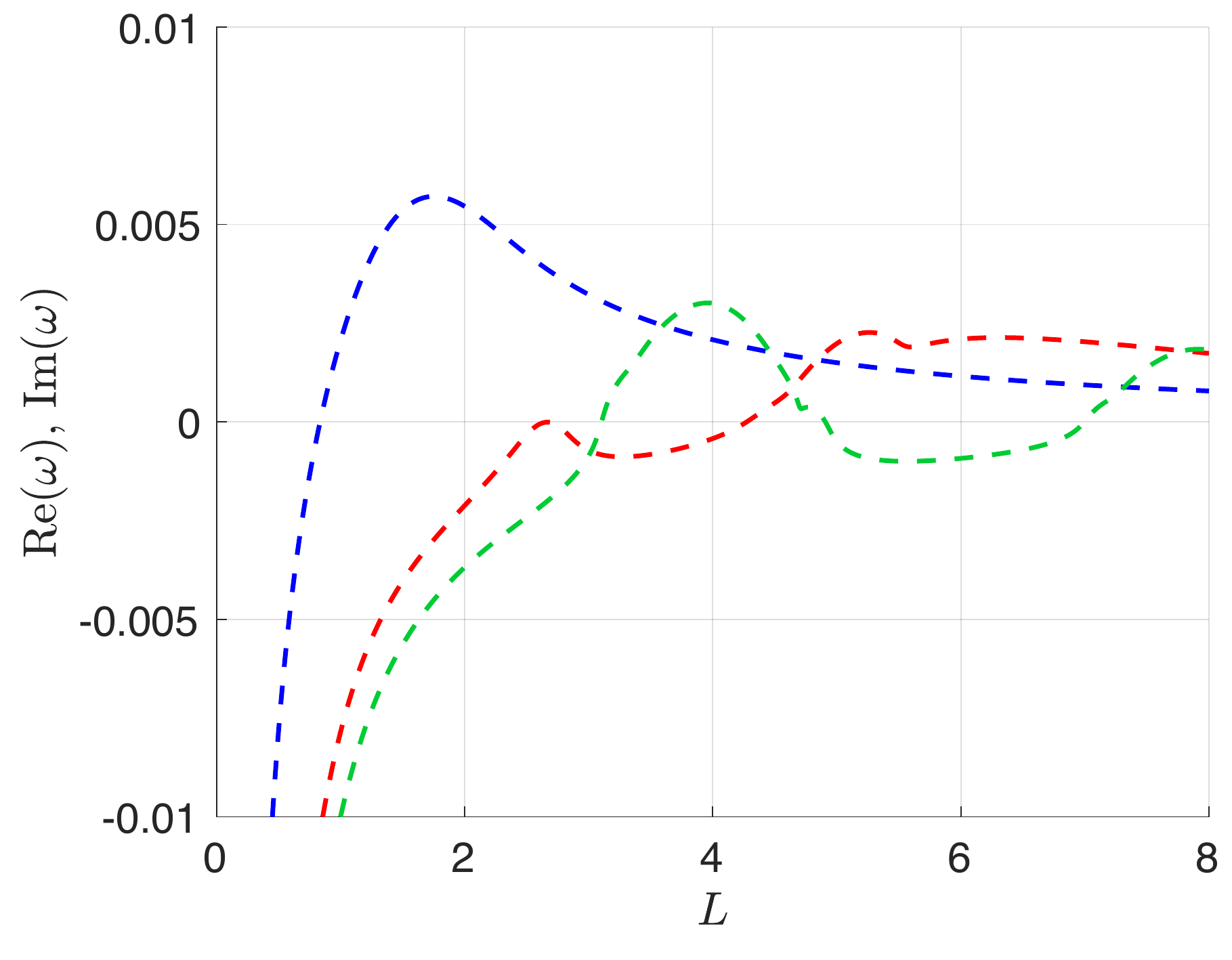}
\caption{Evolution of complex $\om$ as a function of $L$. We have shown the first, third and fifth mode (see remark in Fig.~\ref{BHL_Cplane_Bbc_Fig}). On the right: real parts. On the left: imaginary parts. 
}
\label{BHL_Values_Bbc_Fig} 
\end{figure}

The results for the spectrum and its evolution when varying $L$ are exposed in Figs.~\ref{BHL_Cplane_Bbc_Fig} and \ref{BHL_Values_Bbc_Fig}. The first important difference is that static instabilities are much less frequent. The trajectory of purely imaginary frequency travelling up and down found in the Ingard-Myers case (Fig.~\ref{BHL_Cplane_Fig}) is replaced by a complex frequency coming from the lower half plane, crossing the real line at about $\om \sim \om_\m$, reaching a maximum imaginary part before travelling down to $\om = 0$ (see the blue line in Fig.~\ref{BHL_Cplane_Bbc_Fig}). When this mode is unstable ($\Im(\om) > 0$), it satisfies $\Re(\om) \lesssim \om_\m$. Since $\om_\m$ is a cut-off frequency induced by the boundary layer ($\om_\m = O(\delta)$), such instability is a low frequency one. When the boundary layer thickness is tiny enough, this mode can merge with its partner with respect to \eqref{Sp_Sym} and stick to the imaginary axis, becoming a fully static instability (see Fig.~\ref{BS_Fig}). Upon approaching $\om = 0$, they will split again into a pair with $\Re(\om) \neq 0$. In addition, the maxima reached by the imaginary part of $\om$ are quite smaller. This means that a thin boundary layer tends to \emph{stabilize} the system. 

\begin{figure}[!ht]
\centering
\includegraphics[width=0.7\columnwidth]{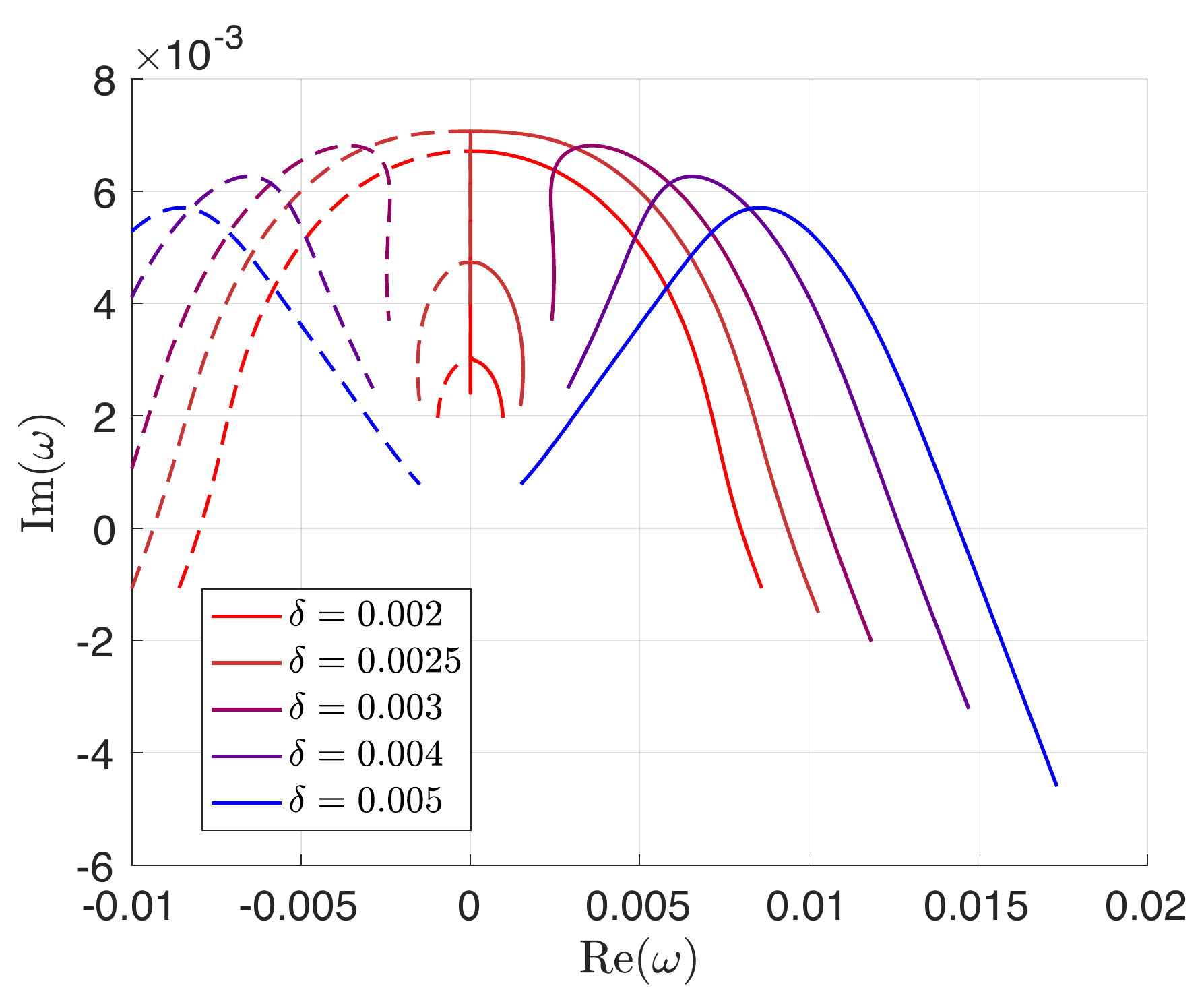}
\caption{Trajectories of the first unstable mode for $L$ from $0.6$ to $3.5$ and for different values of the boundary layer thickness $\delta$. 
}
\label{delta_Plot_Fig} 
\end{figure}

The second main difference is in the region $\Re(\om) > \om_\m$. There is now a series of resonances, with a rather small frequency gap. The explanation is that the boundary layer mode $k_\s$ and the negative energy mode $k_\-$ form a Fabry-Perot like cavity: a wave can travel from right to left with wave number $k_\s$, be converted to a mode of wavenumber $k_\-$ to travel back right. Eigenmodes of this effective cavity are then characterized by having a finite number of wavelength fitting inside the two interfaces (Bohr-Sommerfeld condition). To confirm that this is the case, we compare the real parts of these resonances to the solutions of the condition 
\be \label{BS_eq}
L(k_\s(\om) + k_\-(\om)) = 2n\pi + \nu \qquad (n \in \mathbb Z), 
\ee
where $\nu$ is sum of the phase shifts induced at each interfaces. In the short wavelength limit, $\nu$ is twice the well-known Airy phase shift of $\pi/2$. In general, it stays of order 1. The result is shown in Fig.~\ref{BS_Fig}. We see that the condition captures well the set of Fabry-Perot like resonances. One can notice that the condition predicts a slightly larger frequency gap. This is due to the fact that $\nu$ is not  constant, but slowly varies with $\om$. When varying $L$, the resonances come closer together. From time to time one resonance will come out of the set and migrates to the upper plane, becoming an instability (see trajectories in Fig.~\ref{BHL_Cplane_Bbc_Fig}). This mechanism replaces the appearance of dynamical instabilities ($\Re(\om) \neq 0$) from purely imaginary frequency merging in the Ingard-Myers case. Finally, in Fig.~\ref{BHL_ModesBbc_Fig}, we show the profiles of unstable modes. 

\begin{figure}[!ht]
\centering
\includegraphics[width=0.7\columnwidth]{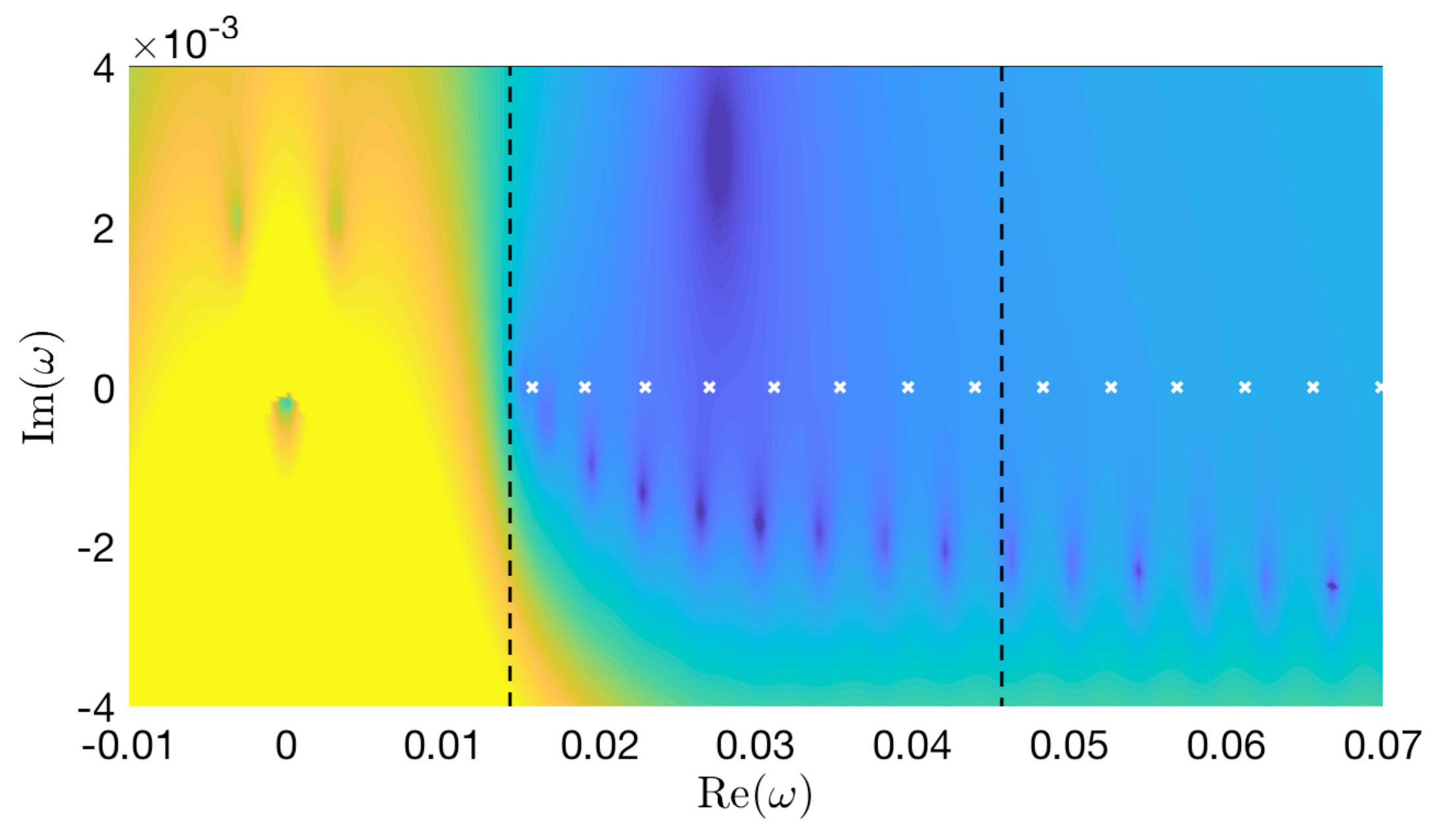}
\caption{Color map of $\ln |\det(a_L,a_B,a_N,a_{A+},a_S)|$ with $M_0=0.3$, $b_C = 6$, $b_O = 15$, $\delta=0.005$ and $L=4$. The white crosses shows the (real) frequencies satisfiying the Bohr-Sommerfeld condition \eqref{BS_eq}, which we have used with $\nu =0$. The dashed black lines mark $\Re(\om) = \om_\m$ and $\Re(\om) = \om_\M$. 
}
\label{BS_Fig} 
\end{figure}

\begin{figure}[!ht]
\centering
\includegraphics[width=0.49\columnwidth]{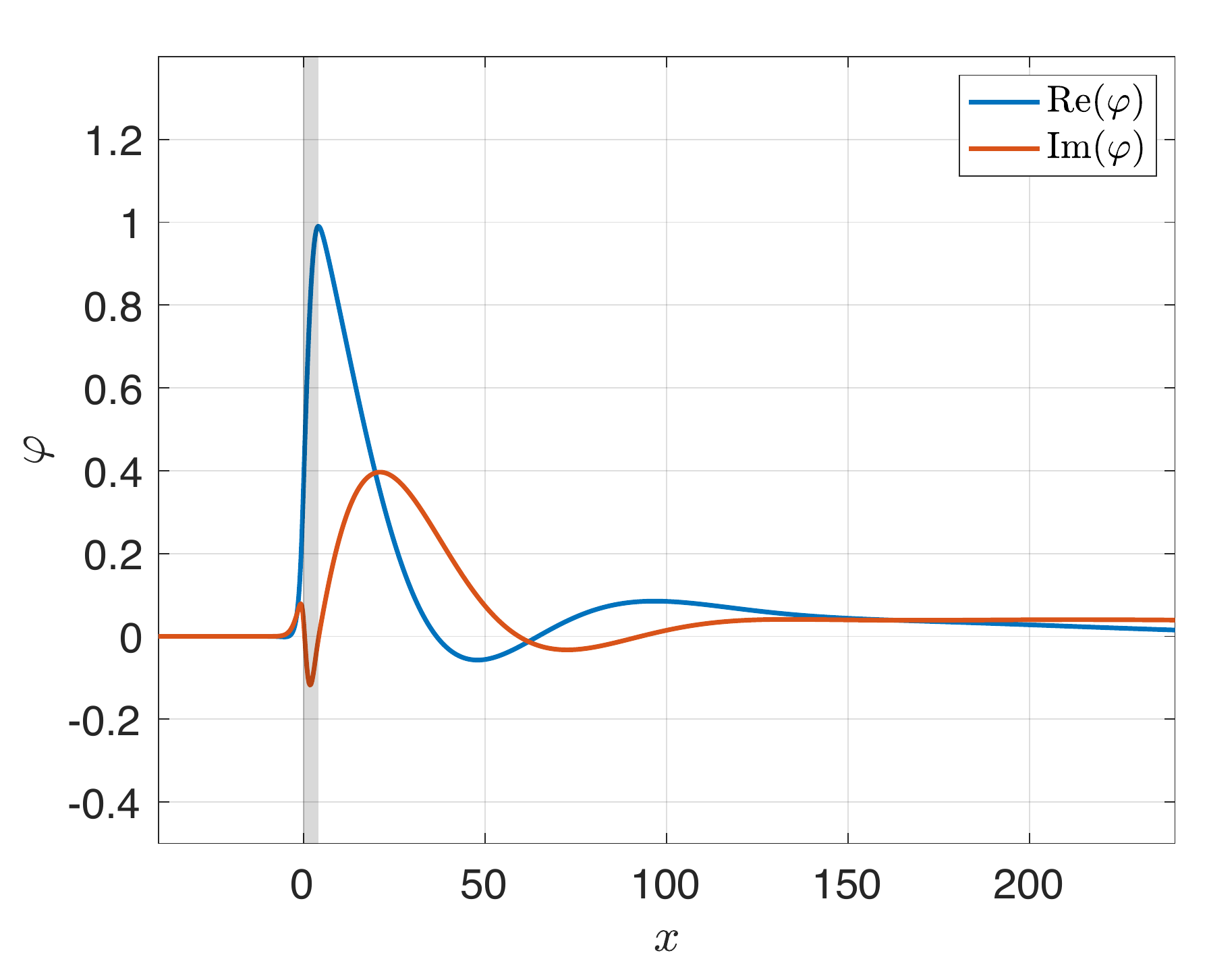}
\includegraphics[width=0.49\columnwidth]{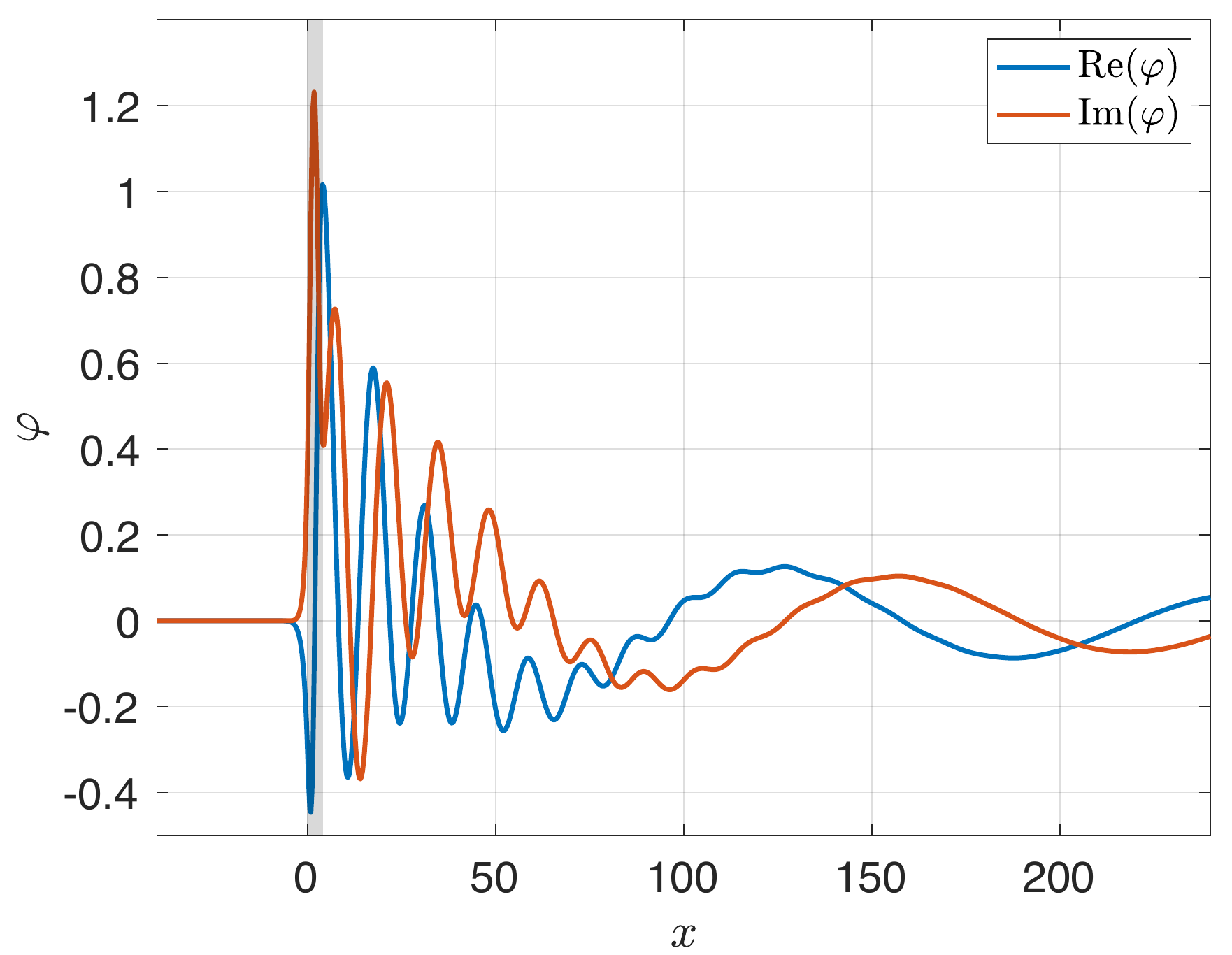}
\caption{Profile of unstable modes with $M_0=0.3$, $b_C = 6$, $b_O = 15$. On the right: first mode for $L=4$ (blue curve in Fig.~\ref{BHL_Cplane_Bbc_Fig}). On the left: fifth mode for $L=4$ (cyan curve in Fig.~\ref{BHL_Cplane_Bbc_Fig}). 
The grey shading delimitates the subsonic region. %CHANGE
}
\label{BHL_ModesBbc_Fig} 
\end{figure}

%%%%%%%%%%%%%%%%%%%%%%%%%%%%%%%%%%%%%%%%%%%%%%%%%%%
%CONCLUSION
%%%%%%%%%%%%%%%%%%%%%%%%%%%%%%%%%%%%%%%%%%%%%%%%%%%

\section{Conclusion}

In this work we have studied a particular configuration of lined ducts, where the flow changes from effectively supersonic ($c_\eff < U_0$) to subsonic ($c_\eff > U_0$) to supersonic again. We have shown that these configurations are subject to a temporal instability similar to the ``black hole laser instability''~\cite{Coutant10}. In addition, we have studied the effect of different boundary conditions taking into account the boundary shear flow layer near the impedance wall: an infinitely small layer (Ingard-Myers condition) and a finite but small layer (modified Brambley condition \eqref{Bram_BC}). 
\bigskip 

This laser instability has several remarkable properties. When using the Ingard-Myers condition, unstable modes divide into two classes: static instabilities, associated with a purely imaginary frequency, and dynamical instabilities with a complex frequency (see Fig.~\ref{BHL_Cplane_Fig}). When taking into account the boundary layer's thickness, we found that static instabilities acquire a finite real part of order $O(\delta)$. Only for very small $\delta$ can one recover static instabilities with $\Re(\om) = 0$ (see Fig.~\ref{delta_Plot_Fig}). Moreover, the presence of a boundary layer tends to the growth rates of unstable modes compared to the Ingard-Myers case. 
\bigskip

The stability of acoustic liner with grazing flow is still much debated, and in particular the role of boundary layers~\cite{Rienstra11,Brambley13}. The present work shows that varying impedances can lead to much stronger instabilities. This is particularly true when transitions from effectively subsonic to supersonic are involved (see Fig.~\ref{BHL_Mplot_Fig}) because they couple together all propagating modes present at a given frequency.

%%%%%%%%%%%%%%%%%%%%%%%%%%%%%%%%%%%%%%%%%%%%%%%%%%%
%APPENDIX
%%%%%%%%%%%%%%%%%%%%%%%%%%%%%%%%%%%%%%%%%%%%%%%%%%%
\newpage
\appendix
\section{Low frequency Brambley condition}
\label{BC_App}
In this appendix, we briefly explain how the effective boundary condition \eqref{Bram_BC} has been obtained from the work of~\cite{Brambley11,Brambley13}. This condition corresponds to the ``short wavelength modified Myers'' derived by Brambley (see section III.B in~\cite{Brambley11} and 4 in~\cite{Brambley13}), but is further simplified by discarding mass, momentum and kinetic energy deficits of the boundary layer (encoded in $I_0$). To derive that boundary condition, it is assumed that the Mach profile, which we denote $M(y)$, is constant equal to $M_0$ except in a region of size $O(\delta)$ near the impedance wall ($\delta$ will be defined more precisely later on). 
The normalized mean density $\rho_0(y)$ is also allowed to vary away from its bulk value in the boundary layer. 
Under this assumption, Brambley has shown that, to first order in $\delta$, the effect of the boundary layer is equivalent to a modified boundary condition: 
\be \label{Full_Bram_BC}
(\p_y \phi)_{y = 1} = i(\om - M_0 k)^2 \frac{1- i\dfrac{\om Z_b k^2}{(\om - M_0 k)^2} I_1(\om; k)}{\om Z_b + i (\om - M_0 k)^2 I_0(\om; k)} \phi_{y = 1}, 
\ee
where $Z_b$ is the impedance of the wall supposed to depend only on the frequency. $I_0$ and $I_1$ are integrals of the boundary layer profiles, given by 
\bsub \bea
I_0(\om; k) &=& \int_0^1 \left[1 - \frac{(\om - M(y) k)^2 \rho_0(y)}{(\om - M_0 k)^2}\right] \d y , \\
I_1(\om; k) &=& \int_0^1 \left[1 - \frac{(\om - M_0 k)^2}{(\om - M(y) k)^2 \rho_0(y)}\right] \d y . 
\eea \esub
Notice that both are of order $O(\delta)$ since $M(y)\sim M_0$ and $\rho_0(y) \sim 1$ outside the boundary layer. We now approximate this boundary condition in the limit of low frequency, more precisely assuming $\om \ll |k|$ \emph{only in the terms of order} $\delta$. The reason is that for small $\delta$ (lower than $1\%$ in this work), the $O(\delta)$ correction of the Ingard-Myers condition is always small unless $\om \ll |k|$. In this limit, while the integral $I_0$ stay bounded, $I_1$ dominates and becomes 
\be \label{I1_eq}
I_1 \sim \frac{\delta M_0 k}{\om}, 
\ee
where the effective boundary layer size $\delta$ (in units of the transverse size of the duct) is given by 
\be
\delta = \frac{M_0}{\rho_{y=1} (M')_{y=1}}. 
\ee
Notice that this expression is exact for a piecewise linear velocity profile. Taking the limit $\om \ll |k|$ in Brambley's condition \eqref{Full_Bram_BC}, as well as the low frequency impedance $Z_b \sim i/(b\om)$, we obtain the effective boundary condition \eqref{Bram_BC} used in the core of the paper. 
\bigskip

At this level, we would like to make a few remarks concerning the validity of our approximate boundary condition \eqref{Bram_BC}. First-of-all, the derivation used by Brambley ignores the possible presence of a critical layer within the boundary layer. This is technically illicit in the range $0 < \om/k < M_0$, which corresponds to the hydrodynamic continuum. Some arguments were presented by Brambley, Darau and Rienstra suggesting that the critical layer could be neglected under reasonable assumptions~\cite{Brambley12}, but the fact that the impedance changes along the wall could on the contrary excite modes in the hydrodynamic continuum~\cite{Dai18}. %A
Moreover, we discussed here only the boundary layer on the side of the impedance wall, but there will be inevitably be another one near the hard wall. This second boundary layer can quite possibly give rise to more propagating modes (as one could see by using again the boundary condition \eqref{Bram_BC} with $b=0$), but we assume here that such extra modes would not couple significantly from the ones present in our model. We believe that a precise understanding of these two effects go beyond the scope of this paper, but it would be interesting to understand how they could affect the laser instability, and more generally the scattering over a spatially varying impedance.
\bigskip

\begin{figure}[!ht]
\centering
\includegraphics[width=0.65\columnwidth]{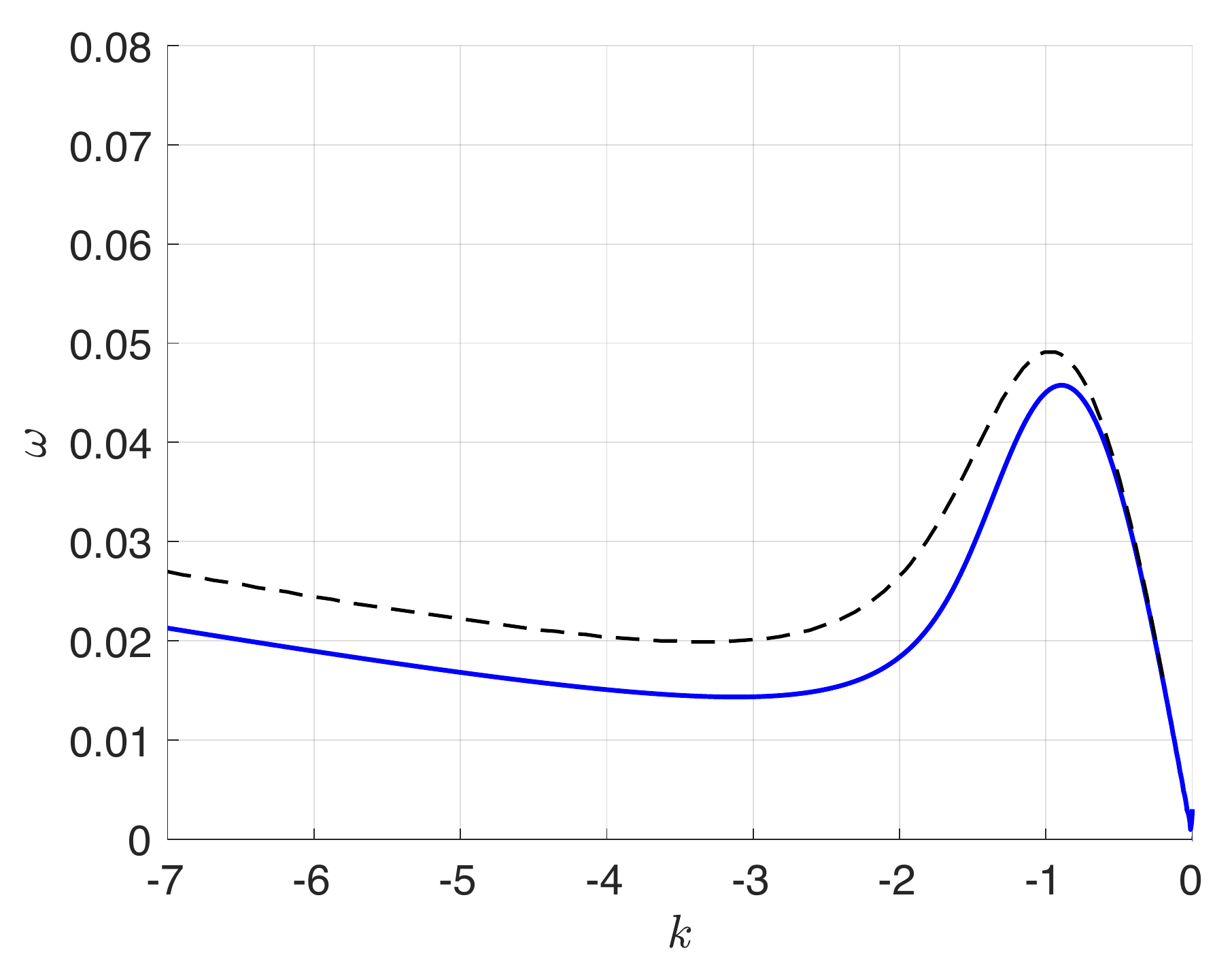}
\caption{Dispersion relation in the $\om > 0$ and $k < 0$ range. The solid blue line is obtained using the one-dimensional model with effective boundary condition (equation \eqref{1D_Disp_rel}), while the dashed black line is obtained by direct integration of the Pridmore-Brown equation. We have used $b=6$, $M_0=0.3$, $\delta=0.005$ and a Mach profile given by $M(y) = M_0(1 - y^{1/\delta})$. 
}
\label{PB_Fig} 
\end{figure}

To gain more confidence in the effective boundary condition used in this work, we compared the dispersion relation obtained with it (equation \eqref{1D_Disp_rel}) with the one obtained by directly solving the Pridmore-Brown equation. In order to avoid complication due to critical layers, we restricted ourselves to $\om > 0$ and $k < 0$. The result is shown in Fig.~\ref{PB_Fig}. Although the differences are quantitatively significant, the qualitative behavior and main features are fairly reproduced. 

\section{Asymptotic solutions of the dispersion relation \eqref{1D_Disp_rel}}
\label{Roots_App}

It is instructive to look at the asymptotics of the solutions of the dispersion relation for small $\delta$. The first thing to notice is that $\om_\m = O(\delta)$, while $\om_\M$ is independent of the boundary layer, i.e. $O(1)$. Thus, in the limit of small $\delta$, there is a separation of scale. Let us consider $\om_\m \ll \om \ll \om_\M$ in a subsonic flow, in which case there are five real solutions. The first four solutions are weakly dependent on $\delta$ (see Fig.~\ref{1DeffDisp_Sub_Fig}) and can be estimated using the Ingard-Myers boundary condition (i.e. $\delta = 0$ in \eqref{Bram_BC}). Moreover, using $\om \ll \om_\M$, we can solve the dispersion relation at first order in $\om/\om_\M$. In that limit, two modes scale as $O(\om)$. These are the acoustic modes, which reads
\bsub \label{1D_ac_roots} \bea
k_\u &\sim& \frac{\sqrt{1+b}}{M_0\sqrt{1+b} - 1} \om , \\
k_\v &\sim& \frac{\sqrt{1+b}}{M_0\sqrt{1+b} + 1} \om . 
\eea \esub
Because of the flow, there are also two hydrodynamic modes $k_\+$, $k_\-$. Using the Ingard-Myers boundary condition, their first order expressions are of the form:  
\bsub \bea
k_\+ &\sim& - k_Z + \frac{\om}{v_g^Z}, \\ 
k_\- &\sim& k_Z + \frac{\om}{v_g^Z} . 
\eea \esub
The value $k_Z$ at zero frequency is simply obtained and reads 
\be
k_Z = \sqrt{\frac{3 - 3(1+b)M_0^2}{b M_0^2(1-M_0^2)}}. 
\ee
A more tedious calculation leads to 
\be
v_g^Z = \frac{M_0(1-M_0^2)(1-M_0^2(1+b))}{1-2M_0^2+M_0^4(1+b)}. 
\ee
The fifth root only exist for finite boundary layer and scales as $1/\delta$. Its asymptotic expression for $\delta \to 0$ is given by 
\be \label{Bram_root}
k_\s \sim -\frac{M_0 b}{\delta} \om. 
\ee
It is also interesting to obtain an asymptotic expression for $\om_\m$ in the limit of small $\delta$. For this, we interpret the fact that $\pm k_Z$ is not a solution of the dispersion relation \eqref{1D_Disp_rel} when $\delta \neq 0$ as an avoided crossing. When $\delta = 0$, the line of $k_\+ \sim - k_Z + \om/v_g^Z$ and $\om = 0$ (the degenerated line of the root $k_\s$ for $\delta = 0$) cross at $k=-k_Z$ and $\om = 0$. Calling $D(\om,k)$ the left-hand side of \eq{1D_Disp_rel}, we approximate the dispersion relation near the crossing by 
\be
\om D(\om,k)_{\delta = 0} \sim \p_k D_Z \om \left(k + k_Z - \frac{\om}{v_g^Z}\right) , 
\ee
where $\p_k D_Z$ is short for $\p_k D(\om = 0, k=-k_Z)$. Now, when $\delta$ is small but non-zero, the dispersion relation near the crossing becomes 
\be
\om D(\om,k) \sim -\p_\om D_Z \left( \om + \frac{\delta k}{Mb} \right) \left( v_g^Z(k + k_Z) - \om\right) - \delta M k_Z^3 \left((1-M^2)k_Z^2 + a\right), 
\ee
where the last term is obtained by taking the limit of $\om D(\om,k)$ at the would-be crossing. Using this we obtain two approximate branches $\om_\pm(k)$. The minimum value of $\om_+$ gives us $\om_\m$. Doing so we have 
\be
\om_\m = \frac{\delta}{M^2 b} \left(\sqrt{\frac{6}{b}} + \sqrt{\frac{3(1-M^2(1+b))}{b(1-M^2)}} \right) + O(\delta^2). 
\ee
Using the same asymptotics, we can also evaluate the norm density \eqref{norm_density} associated with each root of the dispersion relation. Using \eq{Norm_density_PW} in the limit $\delta \to 0$ and $\om/\om_\M \ll 1$, we find 
\bsub \bea
\mathcal E_s[k_{A\pm}] &\sim& (1+b) (\om - M_0 k_{A\pm}) , \\ 
&\sim& (1+b) (\om - M_0 k_{A\pm}) , 
\eea \esub
for the acoustic modes, 
\be
\mathcal E_s[k_{\+/\-}] \sim \mathcal E_s[\mp k_Z] \sim \pm \sqrt{\frac{3 - 3(1+b) M_0^2}{b (1-M_0^2)}} \left(1+ \frac{(1-M_0^2)^2}{b M_0^4}\right), 
\ee
for the hydrodynamic modes, and 
\be
\mathcal E_s[k_\s] \sim \frac{b^5 M_0^6 (1-M_0^2)^2 \om^3}{18 \delta^4} , 
\ee
for the boundary layer mode. We recall that the energy density of each mode is directly obtained by $\mathcal E_e = \om \mathcal E_s$. In these asymptotic expressions, we recover the important result that all modes have a positive energy except for $k_N$, which corresponds to a negative energy wave.

\bibliographystyle{utphys}
\bibliography{Bibli}

\end{document}